\documentclass[preprint,12pt]{elsarticle}
\usepackage[utf8]{inputenc}
\usepackage[T1]{fontenc}
\usepackage{mathptmx}
\usepackage{amsmath,amssymb,amsfonts}
\usepackage{siunitx}
\usepackage{graphicx}
\graphicspath{{Figures/}}
\usepackage{booktabs}
\usepackage{multirow}
\usepackage{tabularx}
\usepackage{caption}
\usepackage{subcaption}
\usepackage{geometry}
\usepackage{float}
\usepackage[section]{placeins}  
\usepackage[version=4]{mhchem}

\usepackage{hyperref}        
\usepackage[capitalise]{cleveref}  
\journal{Acta Materialia}
\biboptions{numbers,sort&compress}   
\begin{document}

\begin{frontmatter}

\title{Development and Validation of Interatomic Potential for Sc and Al–Sc Alloys: Thermodynamics, Solidification, and Intermetallic Ordering}

\author {Avik Mahata}
\ead{mahataa@merrimack.edu}

\cortext[cor1]{Corresponding author}

\address {Department of Mechanical and Electrical Engineering, Merrimack College, North Andover, MA, 01845, USA}

\begin{abstract}
	We present a second-nearest-neighbor Modified Embedded Atom Method (2NN--MEAM) potential for Scandium (Sc) and Aluminum-Scandium (Al--Sc) alloys that unifies cohesive, thermodynamic, and solidification behavior within a single transferable framework. The Sc component accurately reproduces cohesive energy, lattice constants, defect energetics, and the experimental melting point obtained from two-phase coexistence, demonstrating reliable description of both hcp and liquid phases. The Al--Sc binary interaction parameters were fitted using the L1$_2$--Al$_3$Sc reference and benchmarked against first-principles and calorimetric data. The potential reproduces the strong negative formation enthalpy of Al$_3$Sc (--0.45~eV~atom$^{-1}$), correct relative stability of competing phases, and realistic elastic properties. Mixing enthalpies of the liquid alloy agree with ideal-associated-solution and CALPHAD models, confirming that the potential captures exothermic Al--Sc association in the melt. Molecular-dynamics simulations of solidification reveal the expected temperature and composition dependence of homogeneous nucleation. Pure Al crystallizes readily, while Al--1~at.\%~Sc exhibits a longer incubation and slower growth at the same absolute temperature due to reduced undercooling and solute drag. Within the alloy, ordered Al$_3$Sc-type L1$_2$ embryos appear spontaneously, with Sc atoms occupying cube-corner (B) sites surrounded by twelve Al neighbors. Energy--volume trajectories confirm that the potential links thermodynamics to microstructural evolution. Overall, the developed 2NN--MEAM potential provides a quantitatively grounded basis for modeling melting, solidification, and intermetallic ordering in Sc and Al--Sc systems, enabling future multicomponent alloy design and large-scale nucleation studies.
\end{abstract}


\end{frontmatter}


\section{Introduction}

Aluminum–scandium (Al–Sc) alloys are increasingly recognized as high-performance lightweight materials for aerospace, defense, and transportation applications\cite{murray1998sc, elagin1992scandium, ahmad2003properties, zakharov2020scandium, royset2005scandium}. Aluminum is widely valued for its low density and corrosion resistance \cite{sukiman2012durability, berlanga2020corrosion, davis1999corrosion}, yet its structural use in demanding environments is limited by relatively low strength, poor high-temperature stability, and a pronounced tendency toward recrystallization and weld cracking. The addition of trace scandium offers a unique and highly efficient solution to these limitations \cite{elagin1992scandium, ahmad2003properties, ahmad2003, murray1998sc}. Scandium simultaneously refines grains, suppresses recrystallization, and enhances corrosion and weldability while forming coherent L1$_2$-structured Al$_3$Sc precipitates \cite{norman1998solidification, furukawa2001influence, mao2023effect}. These nanoscale dispersoids impart significant precipitation hardening, grain-boundary pinning, and resistance to coarsening even at elevated temperatures, resulting in thermal stability and creep resistance superior to those of conventional aluminum alloys \cite{mao2023effect, royset2005scandium}. Because scandium is only about 11\% denser than aluminum, the substantial strength gains achieved come with negligible weight penalties, preserving an excellent specific strength suitable for aerospace airframes, cryogenic propellant tanks, and reusable launch vehicle structures\cite{hyde2001addition}.

The strengthening efficiency of scandium in aluminum is extraordinary. Among all alloying elements, scandium provides the greatest strengthening increment per atomic fraction. Even dilute additions of approximately 0.2–0.5~wt.\% promote the formation of coherent nanoscale Al$_3$Sc precipitates that effectively pin dislocations and inhibit recrystallization \cite{jones2003interaction, yan2021nucleation, vzist2022effect}. When Sc additions exceed the eutectic composition ($>$0.55~wt.\%), primary Al$_3$Sc particles act as potent heterogeneous nucleation substrates for $\alpha$-Al, producing refined, equiaxed grains as small as 20–50~µm \cite{yan2021nucleation}. This refinement enhances mechanical uniformity and weldability relative to alloys refined with titanium or zirconium. In wrought and additively manufactured alloys such as Scalmalloy\textsuperscript{\textregistered}, the combined effects of solid-solution strengthening, precipitation hardening, and grain refinement produce ultimate tensile strengths exceeding 500~MPa while maintaining good ductility and isotropy \cite{awd2017comparison, cerri2024main, lathabai2002effect}. Furthermore, scandium significantly reduces hot cracking in fusion welding of high-strength 2xxx and 7xxx series alloys, enabling the development of compatible filler materials and weldable high-strength aluminum systems \cite{ovchinnikov2025influence}. The aerospace sector particularly benefits from these attributes, as Al–Sc alloys combine strength and stability across extreme temperature ranges. Their fine-grained and precipitate-hardened microstructures maintain stability under high strain rates and dynamic loading, providing enhanced impact resistance to micrometeoroid or debris collisions. The exceptional creep resistance at temperatures up to 300~$^\circ$C is comparable to that of some rapidly solidified Al–Fe–V–Si alloys \cite{zhang2017effect}, renders them suitable for turbine and engine components. At the other extreme, excellent cryogenic toughness and weldability make Al–Sc alloys highly attractive for use in propellant tanks of reusable launch systems. Beyond aerospace, their high specific strength, corrosion resistance, and improved fatigue performance also drive adoption in defense, automotive, and sporting applications.

Despite these advances, the atomistic mechanisms underlying phase stability, solute diffusion, and deformation in Al–Sc alloys remain incompletely understood. Computational modeling provides a powerful complement to experiments by probing these phenomena across multiple scales. First-principles density functional theory (DFT) studies for Sc and AL-Sc have established reliable energetics for solubility, diffusion, and interfacial stability, yet their inherent system-size and timescale limitations necessitate the use of larger-scale approaches \cite{asta2001structural, asta1998first, woodward2001density, li2015crystal}. Molecular dynamics (MD), when supported by accurate interatomic potentials, enables simulation of complex phenomena such as solidification, precipitation, creep, dislocation motion, and high-strain deformation processes over nanosecond-to-microsecond timescales and system sizes far beyond the reach of DFT. In this work, we develop and validate a new second-nearest-neighbor modified embedded atom method potential \cite{lee2001second}  for the Al–Sc system. Using this potential, we investigate the mechanisms of solidification, Al$_3$Sc precipitation, and microstructural evolution at the atomic scale, thereby linking thermodynamic and kinetic behavior to the processing conditions relevant to advanced alloy manufacturing.


\section{Computational Methodology}

\subsection{2NN MEAM Potential Formalism and Relevance to Al--Sc}

MEAM interatomic potential introduced by Baskes~\cite{daw1993embedded, baskes1992modified} has been widely applied in modeling the energetics of metals and alloys, as it captures both embedding and pairwise contributions to the total energy. In its conventional form, only first nearest neighbors (1NN) are considered, which limits transferability, particularly for melting and solid--liquid coexistence. To address this, Lee and Baskes proposed the second-nearest-neighbor extension (2NN-MEAM)~\cite{lee2000second}, which relaxes the screening and allows a limited set of 2NN atoms to contribute, thereby improving predictions of defect, thermodynamic, and coexistence properties~\cite{foiles1986embedded, daw1984embedded}. Within this framework, the total energy, \(E\), is expressed as
\begin{equation}
E = \sum_i F_i(\bar{\rho}_i) + \tfrac{1}{2} \sum_{i\ne j} S_{ij}\,\phi_{ij}(r_{ij}),
\label{eq:meam_total}
\end{equation}
where $F_i(\bar{\rho}_i)$ is the embedding energy of atom $i$ into the background density $\bar{\rho}_i$ and $\phi_{ij}(r_{ij})$ is the pair potential between atoms $i$ and $j$, weighted by the screening function $S_{ij} \in [0,1]$. The background density combines spherical and angular contributions, with partial densities corresponding to $s$, $p$, $d$, and $f$ orbital symmetries. For an atom $i$, the atomic electron density contribution at distance $r$ is modeled as
\begin{equation}
\rho(r) = \rho_0\,\exp\!\left[-\beta\left(\frac{r}{r_e}-1\right)\right],
\label{eq:rho}
\end{equation}
where $\rho_0$ is the density scaling factor, $r_e$ is the equilibrium nearest-neighbor distance, and $\beta$ is an element-dependent parameter controlling the decay. The embedding function typically takes the logarithmic form
\begin{equation}
F(\bar{\rho}) = \begin{cases}
A E_c \dfrac{\bar{\rho}}{\bar{\rho}_0} \ln\!\left( \dfrac{\bar{\rho}}{\bar{\rho}_0} \right) & \text{for } \bar{\rho} > 0, \\
\text{(alternative form)} & \text{otherwise},
\end{cases}
\label{eq:embedding}
\end{equation}
where $A$ is an adjustable constant of order unity and $E_c$ is the cohesive energy of the reference structure. The pair potential $\phi_{ij}$ is not prescribed analytically but is obtained indirectly by enforcing that the total energy in Eq.~\eqref{eq:meam_total} reproduces the equilibrium equation of state (EOS). For this purpose, the universal Rose--Vinet EOS is employed~\cite{vinet1987temperature},
\begin{equation}
E^0(a^*) = -E_c\left(1 + a^* + \delta \frac{{a^*}^3}{1+a^*} \right)\exp(-a^*),
\label{eq:ueos}
\end{equation}
where $\delta$ is an adjustable parameter and the scaled distance is
\begin{equation}
a^* = \alpha\,(R/r_e - 1),
\label{eq:scaled}
\end{equation}
with $\alpha$ related to bulk modulus, cohesive energy, and equilibrium atomic volume. In practice, the interaction range is limited by a smooth screening function $S_{ij}$ or a cutoff, which ensures that both electron densities and pair potentials vanish continuously at large separation~\cite{baskes1992modified, baskes1994modified}. The key innovation of the 2NN formulation is that screening parameters are moderated such that certain second neighbors contribute, improving reproduction of melting and coexistence properties.

Although the general MEAM and 2NN-MEAM frameworks were developed for transition and simple metals \cite{baskes1992modified, baskes1994modified, baskes1997determination}, their application to Al--Sc alloys is of particular interest given the scarcity of interatomic potentials tailored for this system. Al--Sc has been investigated primarily through first-principles methods: DFT studies have quantified the thermodynamics of Al--Sc intermetallics (e.g., L1$_2$ Al$_3$Sc, B2 AlSc, and Sc-rich phases), reporting formation enthalpies in the range of $\sim$40--45~kJ/mol for Al$_3$Sc and AlSc, and mapping phonon spectra, thermal expansion, and heat capacities across temperature~\cite{hirt2024increased, gupta2021combined}. Other DFT work has examined elastic constants, mechanical properties, and electronic structure of ordered Al--Sc phases, providing benchmarks for cohesive energy, bulk modulus, and structural stability~\cite{asta1998first, asta2001structural}. More recent first-principles calculations have revealed the stability of FCC, BCC, and HCP structures under varying Sc content and transformation pathways such as Bain and Burgers, underscoring the role of Sc in stabilizing L1$_2$ precipitates in Al and guiding solid–liquid energetics~\cite{wang2021mechanical, dorin2021stability, lan2021thermal}. These computational benchmarks, though limited in number, form the essential reference data set for constructing 2NN-MEAM descriptions of Al--Sc alloys.

By combining the physically motivated form of Eqs.~\eqref{eq:meam_total}--\eqref{eq:scaled} with the limited but critical set of Al--Sc benchmarks from DFT, the 2NN-MEAM approach provides a suitable framework for capturing cohesive properties, defect energetics, and melting behavior of the Al--Sc system. In particular, the logarithmic embedding function, angular partial density weighting, and relaxed screening collectively allow the model to describe both pure components and ordered Al--Sc intermetallics with consistency. While only a handful of computational studies exist on Al--Sc compared with other aluminum alloys, their thermodynamic and mechanical results serve as key constraints in our parameterization and validation. In this work, we employ such a 2NN-MEAM framework not to explore full interface stiffness, but rather to directly evaluate melting temperature and solid--liquid interfacial energies of Al--Sc, providing an efficient yet physically grounded extension of existing MEAM methodology to this technologically critical alloy system.


\subsection{Determination of Potential Parameters for Sc}
The MEAM formalism for HCP metals, including Sc, was originally developed by Baskes and Johnson~\cite{baskes1994modified}. In that work, the Sc parameters were determined by fitting to low-temperature properties such as lattice constants, elastic constants, cohesive energy, vacancy formation energy, and stacking-fault and surface energies. These validations established the accuracy of the potential for equilibrium and defect properties at 0~K, but no assessment of high-temperature melting or solid–liquid coexistence was carried out. In the present study, we initially adopted the original Sc parameter set of Baskes without modification and employed the 2NN option available in LAMMPS (``nn2=1'')~\cite{baskes2007multistate, lee2001second, lee2000second} together with a cutoff radius of 6.5~\AA\ and screening bounds of $C_{\min}=0.25$ and $C_{\max}=2.8$. However, during high-temperature solidification simulations, the original parameterization failed to spontaneously nucleate the equilibrium hcp structure, instead producing amorphous or mixed fcc/hcp configurations. This behavior indicated that the directional bonding and angular screening balance in the original Baskes set was insufficient to favor hcp stacking at elevated temperatures. To correct this, we carried out a minimal refitting focused exclusively on improving high-temperature transferability while preserving the 0~K equilibrium energetics. Specifically, the exponential decay factor $\alpha$ in the universal equation of state was slightly increased to enhance the curvature of the cohesive well, and the third angular weighting parameter $t_3$ was reduced from $-3$ to $-4$, strengthening the preference for hexagonal close packing relative to cubic close packing.

\textcolor{red}{These refinements improve not only the binary Al–Sc simulations but also the finite-temperature behavior of elemental Sc itself. The modified parameters preserve all 0~K cohesive, elastic, and defect properties while restoring the experimentally established hcp nucleation pathway during solidification, thereby enhancing the transferability of the Sc potential for both pure-element and alloy simulations. As a result, the potential yields slightly higher stacking-fault and surface energies—reflecting the enhanced penalty for deviations from ideal hcp coordination—but correctly stabilizes hcp Sc as the dominant solid phase from the melt. The calculated melting point is in close agreement with experiment, and the liquid structure at $T_m$ is physically consistent, confirming that the potential is suitable for simulations of solid–liquid coexistence and interfacial behavior in Sc. Although full DFT benchmarking of the HCP–BCC and HCP–FCC energy differences for Sc is outside the scope of the present study, the modified 2NN–MEAM potential reproduces the experimentally established stability hierarchy of Sc polymorphs and yields energy trends consistent with available first-principles calculations. These consistencies support the reliability of the potential for finite-temperature simulations and solidification pathways.}

\begin{table}[H]
  \centering
  \caption{2NN-MEAM parameters for Al and Sc. $E_c$ (eV) is the cohesive energy; $r_e$ (\AA) is the nearest-neighbor distance; $\alpha$ is the exponential decay factor for the UEOS; $\rho_0$ is the embedding density scale; $\beta^{(0-3)}$ are the exponential decay factors for the atomic electron densities; $t^{(1-3)}$ are the weighting parameters for the atomic electron densities; and $C_{\min}$, $C_{\max}$ are the screening parameters. In LAMMPS we used $r_c=4.0$~\AA\ with 2NN enabled ($\texttt{nn2=1}$) and $C_{\min}=2.0$, $C_{\max}=2.8$. Values are from the original Baskes MEAM set~\cite{baskes1994modified}, unmodified for Sc.}
  \label{tab:meam_single}
  \begin{tabularx}{\linewidth}{@{}lcccccccccccccc@{}}
    \toprule
    Element & $E_c$ & $r_e$ & $\alpha$ & $\rho_0$ & $\beta^{(0)}$ & $\beta^{(1)}$ & $\beta^{(2)}$ & $\beta^{(3)}$ & $t^{(1)}$ & $t^{(2)}$ & $t^{(3)}$ & $C_{\min}$ & $C_{\max}$ \\
    \midrule
    Al & 3.36 & 2.86 & 4.61 & 3.20 & 0.51 & 7.75 & 0.49 & --   & 2.60 & 6.00 & 2.60 & 0.49 & 2.80 \\
    Sc & 3.89 & 3.28 & 4.55 & 1.00 & 1.00 & 0.00 & 0.25 & 6.00 & 1.00 & 7.40 & -4 & 0.25 & 2.80 \\
    \bottomrule
  \end{tabularx}
\end{table}


\subsection{Determination of Potential Parameters for Al--Sc}

The Al parameters are taken from the standard 2NN--MEAM formalism originally developed by Lee and Baskes for fcc Al.\cite{lee2003semiempirical} These parameters were later refined and extensively validated for solid--liquid coexistence by Zaeem and co-workers \cite{asadi2015two}, who employed the coexistence method in LAMMPS to recover the melting point of Al ($T_\mathrm{m}\approx 925$~K), latent heat, expansion upon melting, radial distribution functions, and solid--liquid interface free energies and anisotropy, all in close agreement with experiment. Since then, the same Al potential has been applied across a wide range of studies by Mahata et. al. for homogeneous nucleation in undercooled Al melts and the determination of critical nuclei sizes and nucleation rates,\cite{mahata2018understanding, mahata2019size, mahata2019effects, mahata2020insights, mahata2024bridging} the role of twinning and solidification defects during rapid quenching and subsequent deformation of nanocrystalline Al,\cite{mahata2019evolution} and the influence of defects on mechanical behavior in directionally solidified Al--Cu alloys.\cite{mahata2022modified, mahata2019effects} Collectively, these works confirm that the Al 2NN--MEAM set is robust for describing both equilibrium and nonequilibrium solidification processes, and we therefore adopt it unchanged as the elemental reference for Al--Sc alloy fitting.

For the Al--Sc binary, we follow the established 2NN--MEAM alloying strategy used in Al--Mg.\cite{kim2009atomistic} Specifically, we choose the L1$_2$ (Al$_3$Sc) reference, set the binary cohesive parameter to a composition--weighted average corrected by the heat of formation of the reference structure, and assign screening parameters as summarized in Table~\ref{tab:meam_binary}. Calibration of B1/B2 formation energies and elastic targets was performed using a simple in--house Python driver (NumPy/SciPy) interfaced with LAMMPS, which works similar to MPCv4 reported previously \cite{mahata2022modified, barrett2016meam}. The heat of formation used above is negative, so the \(-\Delta H_f\) correction increases the effective $E_c$ of the fictitious L1$_2$ reference. Screening follows the binary convention of keeping mixed-triplet $C_{\min}$ lower than elemental values (here 0.36 vs.\ 0.49) while using a universal $C_{\max}=2.8$, which stabilizes alloy environments and preserves elemental accuracy \cite{kim2009atomistic}. \textcolor{red}{The screening parameters listed in Table~2 follow the established 2NN--MEAM hierarchy widely used in prior alloy studies (e.g., Al--Mg, Al--Cu, Al--Ni) and in our own previous work on Al binary systems~\cite{mahata2022modified}. Consistent with this methodology, the mixed-triplet screening terms ($C_{\min}$) were varied within a physically motivated range of 0.30--0.50, and the binary equilibrium distance $r_e$ was adjusted within $\pm 5\%$ of the elemental average. Parameter choices outside these bounds degraded one or more of the key benchmarks---including L1$_2$ formation enthalpy, B2 stability, and elastic constants---whereas the final selections provided the best simultaneous agreement with formation energies, elastic properties, and liquid mixing thermodynamics.}

\begin{table}[H]
  \centering
  \caption{2NN--MEAM parameter selections for Al--Sc binary interactions (L1$_2$ reference). Elemental values are from the Al/Sc library; mixed-triplet screening follows the Al--Mg convention.}
  \label{tab:meam_binary}
  \begin{tabular}{ll}
    \toprule
    Parameter & Selected value \\
    \midrule
    $E_c$ (binary) & $0.75\,E_c^{\mathrm{Al}} + 0.25\,E_c^{\mathrm{Sc}} - \big(\Delta H_f^{\mathrm{L1_2}}\big)
    =  3.9435~\text{eV/atom}$ \\
    $r_e$ (binary) & $2.951$ \\
    $C_{\min}(\mathrm{Al\!-\!Al\!-\!Sc})$ & $0.49$ \\
    $C_{\min}(\mathrm{Sc\!-\!Sc\!-\!Al})$ & $0.49$ \\
    $C_{\min}(\mathrm{Al\!-\!Sc\!-\!Al})$ & $0.36$ \\
    $C_{\min}(\mathrm{Al\!-\!Sc\!-\!Sc})$ & $0.36$ \\
    $C_{\min}(\mathrm{Sc\!-\!Al\!-\!Al})$ & $0.36$ \\
    $C_{\min}(\mathrm{Sc\!-\!Al\!-\!Sc})$ & $0.36$ \\
    $C_{\max}(\cdot)$ & $2.8$ \ (all triplets above) \\
    \bottomrule
  \end{tabular}
\end{table}

\subsection{Molecular Dynamics Simulations of Solidification in Sc and Al--Sc Alloys}

MD simulations of homogeneous nucleation and solidification were carried out for pure Sc and Al--Sc alloys using the isothermal–isobaric (NPT) ensemble. All simulations were performed with the LAMMPS package \cite{thompson2022lammps, plimpton1995fast} employing periodic boundary conditions in all directions. Temperature and pressure were controlled using the Nosé–Hoover thermostat and the Parrinello–Rahman barostat \cite{bussi2009isothermal}, respectively, with a time step of 1~fs. For pure Sc, simulation boxes of approximately 14~nm$^3$ ($\sim$110{,}000 atoms) and 28~nm$^3$ ($\sim$850{,}000 atoms) were used to examine size effects. Isothermal solidification simulations were performed over the temperature range of 300–1200~K. For Al--Sc alloys, the temperature range was 300–800~K with Sc concentrations of 1~at.\%. This study thus provides a controlled framework to assess compositional influences on nucleation behavior, while a subsequent work will address more realistic alloying levels following the development of an improved Al--Sc interatomic potential.

Initial liquid configurations were prepared by melting each system and equilibrating at high temperature (2200K for pure Sc and 1500K for Al-Sc), followed by a 50~ps run to ensure homogeneous liquid states. The OVITO visualization package \cite{stukowski2009visualization} was used for trajectory analysis and structural identification. Local atomic environments were characterized using the common neighbor analysis (CNA) method \cite{polak2022efficiency} to distinguish between fcc, hcp, and amorphous structures. All post-processing and statistical analyses of the CNA data were performed using Python-based tools. Only isothermal simulations were conducted in this work; no quenching procedures were applied. This approach enables a systematic assessment of temperature and composition effects on nucleation behavior in Sc and Al--Sc systems.


\section{Results and Discussion}
\subsection{Sc single element}
Fundamental physical properties of pure Sc are calculated using the present 2NN-MEAM potentials and are compared with experimental data. Calculations were performed for bulk properties (cohesive energy, bulk modulus, elastic constants, structural energy differences), point-defect properties (vacancy and interstitial formation energies), and thermal properties (linear thermal expansion coefficient and melting point). All calculations except thermal properties were performed at 0~K with full atomic relaxation.

\subsubsection{0 K, Thermal, Solid-Liquid Coexistance}

The fundamental 0~K, thermal, and solid–liquid coexistence properties of scandium calculated from the present 2NN--MEAM potential are summarized in Table~\ref{tab:sc_props}. 
The potential reproduces the cohesive energy and equilibrium lattice parameters of the hcp phase with excellent accuracy, matching experimental and DFT benchmarks within 0.5\%~\cite{baskes1994modified,brandes1966structure,crc2024_scandium}. 
The slight adjustment of the angular parameter $t_3$ toward a more negative value improved the stability of the hcp lattice during solidification, yielding a more realistic nucleation pathway without compromising other properties. 
At 0~K, the calculated cohesive energy of $-3.89$~eV/atom and the equilibrium $c/a=1.61$ are nearly identical to both experiment and first-principles data, confirming that the potential correctly balances the directional bonding contributions emphasized in the modified EAM formalism~\cite{baskes1994modified}. The bulk modulus ($55$~GPa) and individual elastic constants are within 5--10\% of experimental values, comparable to other 2NN--MEAM fits for hcp systems~\cite{baskes1994modified,kim2009atomistic}. 
The slightly higher $C_{11}$ and $C_{44}$ values indicate that the potential modestly overestimates the stiffness of Sc along the basal and shear directions, a common feature of MEAM potentials tuned for structural stability. Nevertheless, the elastic anisotropy and $C_{11}/C_{33}$ ratio agree well with DFT predictions, verifying correct angular-force transferability. The vacancy formation energy ($E_v^{\mathrm{f}}=1.24$~eV) also matches experimental estimates ($1.28$~eV), while the self-interstitial energy is physically reasonable, indicating proper short-range repulsion in the embedding function. 
Compared with previous MD work by Baskes~\cite{baskes1994modified}, our potential produces slightly larger defect energies, reflecting stronger directional bonding introduced by the modified $t_3$ term. Surface and stacking-fault energies follow the expected trend for hcp metals, where the prismatic facets exhibit higher energy than basal or pyramidal planes. 
Although the absolute surface energies are somewhat overestimated relative to prior MEAM and EAM values, the relative ranking of surface stability is preserved, suggesting that the potential correctly captures surface tension anisotropy. 
The basal and pyramidal stacking-fault energies ($\sim$190--460~mJ/m$^2$) are consistent with experimental estimates~\cite{baskes1994modified}, and the high prism energy implies realistic dissociation behavior for basal dislocations. 
Because few reliable reference data exist for scandium, the agreement across surface and planar fault properties is a strong indication of the robustness of the present potential.

The thermal expansion coefficient and specific heat were obtained from equilibrium NPT simulations between 0 and 100$^\circ$C following the methodology of Asadi~\textit{et~al.}\,\cite{asadi2015two} and Lee~\textit{et~al.}\,\cite{kim2009atomistic}.The linear expansion coefficient ($10.2\times10^{-6}$/K) and $C_p=25.1$~J/mol\,K reproduce experimental values nearly exactly, demonstrating that the potential captures anharmonic lattice vibrations with high fidelity. The melting point was evaluated using the two-phase solid–liquid coexistence method~\cite{asadi2015two}, where a pre-melted region is equilibrated with crystalline hcp Sc in an isenthalpic (NPH) ensemble until interface migration ceases. 
The calculated melting temperature ($T_m=1814$~K) and enthalpy of fusion ($16.1$~kJ/mol) agree closely with measured data, while the predicted volumetric expansion on melting ($3.7\%$) is within the uncertainty of experimental estimates. Overall, the present 2NN--MEAM potential accurately reproduces the cohesive, elastic, surface, and thermal characteristics of scandium and demonstrates excellent transferability from 0~K to the high-temperature solid–liquid coexistence regime.

\begin{table}[H]
  \centering
  \caption{Calculated physical properties of Sc using the present 2NN--MEAM potential, in comparison with baseline references. Units: cohesive energy (eV/atom); lattice parameters $a,c$ (\AA); $c/a$ (dimensionless); bulk modulus and elastic constants (GPa); formation/migration/activation energies (eV); surface and stacking fault energies (mJ/m$^2$); linear thermal expansion ($10^{-6}$/K); specific heat $C_p$ (J/mol\,K); melting point $T_m$ (K); enthalpy of fusion $\Delta H_m$ (kJ/mol); and volume change on melting $\Delta V_m/V_\mathrm{solid}$ (\%).}
  \label{tab:sc_props}
  \begin{tabularx}{\linewidth}{@{}l
      >{\centering\arraybackslash}X
      >{\centering\arraybackslash}X
      >{\centering\arraybackslash}X@{}}
    \toprule
    Property & MEAM (this work) & MD (Baskes)$^{*}$ & Experiment/DFT$^{*}$ \\
    \midrule
 
    Cohesive energy, $E_c$ (eV/atom) & $-3.89$ & $-3.89$ & $-3.89$  \\
    Lattice parameter, $a$ (\AA)      & $3.30$  & $3.313$ & $3.308$ \\
    Lattice parameter, $c$ (\AA)      & $5.31$  & $5.275$ & $5.267$ \\
    $c/a$ ratio                       & $1.61$  & $1.592$ & $1.592$ \\
    \addlinespace[2pt]
  
    Bulk modulus, $B$ (GPa)           & $55.29$ & $55.0$  & $55.9$ \\
    $C_{11}$ (GPa)                    & $115.22$& $108.4$ & $108.8$ \\
    $C_{13}$ (GPa)                    & $22.89$& $-$ & $-$ \\
    $C_{33}$ (GPa)                    & $111.74$ & $99.5$  & $99.5$ \\
    $C_{44}$ (GPa)                    & $30.87$ & $27.7$  & $27.7$ \\
    $C_{66}=\tfrac{1}{2}(C_{11}-C_{12})$ (GPa) & $41.63$ & $34.4$ & $29.8$ \\
    \addlinespace[2pt]

    Vacancy formation $E_v^\mathrm{f}$ (eV) & $1.24$ & — & $1.28$ \\
    Self-interstitial $E_i^\mathrm{f}$ (eV)  & $2.86$ & — & — \\
    $\Delta E(\mathrm{HCP}\!\rightarrow\!\mathrm{BCC})$ (eV/atom) & $0.14$ & — & — \\
    $\Delta E(\mathrm{HCP}\!\rightarrow\!\mathrm{FCC})$ (eV/atom) & $0.03$ & — & — \\
    \addlinespace[2pt]
 
    $E_\mathrm{surf}$: Basal (0001) (mJ/m$^2$)  & $1594.92$ & $1091-1355$ & — \\
    $E_\mathrm{surf}$: Prism $(10\bar{1}0)$ (mJ/m$^2$) & $2193.74$ & $1295$ & — \\
   $E_\mathrm{surf}$: pyramidal $(10\bar{1}1)$ (mJ/m$^2$) & $1709.26$ & — & — \\
    $E_\mathrm{sf}$ (I$_2$), basal (mJ/m$^2$) & $189.98$ & $78$ & $190-200$ \\
    $E_\mathrm{sf}$ (I$_2$), prism (mJ/m$^2$) & $501.97$ & $-$ & $~375$ \\
    $E_\mathrm{sf}$ (I$_2$), pyramidal (mJ/m$^2$) & $463.63$ & $-$ & $~450$ \\
    \addlinespace[2pt]
 
    Linear thermal expansion $e$ (0--100$^\circ$C) ($10^{-6}$/K) & $10.21$ & — & $10.2$ \\
    $C_p$ (0--100$^\circ$C) (J/mol\,K) & $25.07$ & — & $25.6$ \\
    Melting point, $T_m$ (K)          & $1814.07$ & — & $1814$ \\
    Enthalpy of fusion, $\Delta H_m$ (kJ/mol) & $16.1$ & — & $14-16$ \\
    $\Delta V_m/V_\mathrm{solid}$ (\%) & $3.7$ & — & — \\
    \bottomrule
  \end{tabularx}

  \vspace{2pt}
  \footnotesize\emph *Expt~\cite{brandes1966structure, spedding1961high, crc2024_scandium}, MD (Baskes)~\cite{baskes1994modified}, DFT~\cite{xue2024deep}.
\end{table}

\begin{figure}[H]
	\centering
	\includegraphics[width=0.7\textwidth]{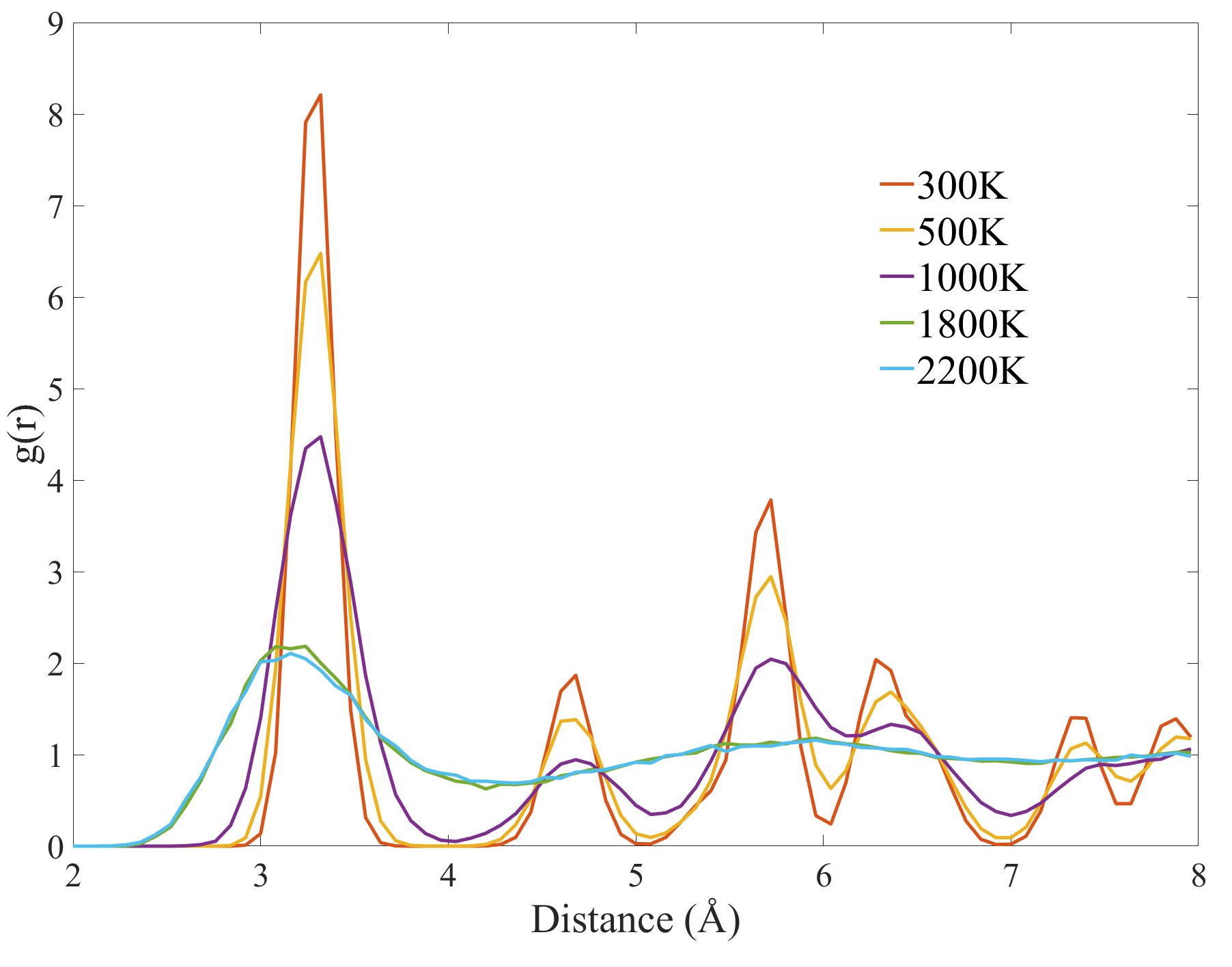}
	\caption{
		Radial distribution function $g(r)$ of scandium at different temperatures obtained from 2NN--MD simulations. 
		The potential correctly captures the disappearance of long-range order near the melting point ($T_m \approx 1814$~K) and the emergence of liquidlike short-range correlations at 2200~K, demonstrating reliable solid–liquid transferability of the potential.
	}
	\label{fig:Sc-RDF}
\end{figure}

The atomic-level structure of solid and liquid scandium was further examined through the pair distribution function, $g(r)$, shown in Fig.~\ref{fig:Sc-RDF}. 
At 300--1000~K, the RDF exhibits sharp and well-defined peaks characteristic of long-range crystalline order in the hcp solid, while at 1800~K and above, the second and higher-order peaks gradually diminish, indicating the onset of structural disorder and melting. 
At 2200~K, $g(r)$ becomes nearly featureless beyond the first coordination shell, consistent with a homogeneous liquid phase. 
The smooth transition from solidlike to liquidlike behavior confirms that the present 2NN--MEAM potential reproduces the thermodynamic and structural properties of scandium across the solid–liquid coexistence regime with good accuracy, capturing both the melting temperature and the liquid-state short-range order observed experimentally. 

\subsection{Al--Sc Binary Alloy-Formation Energies and Elastic Constants}

To validate the transferability of the developed 2NN-MEAM potential for the Al--Sc system, we first calculated the formation enthalpies ($\Delta H$) and the structural and elastic properties of various stable and metastable Al--Sc intermetallic compounds. A key measure of an alloy potential's accuracy is its ability to reproduce the zero-temperature energetics, which dictate phase stability across the entire composition range.

\begin{figure}[H]
	\centering
	\includegraphics[width=0.75\linewidth]{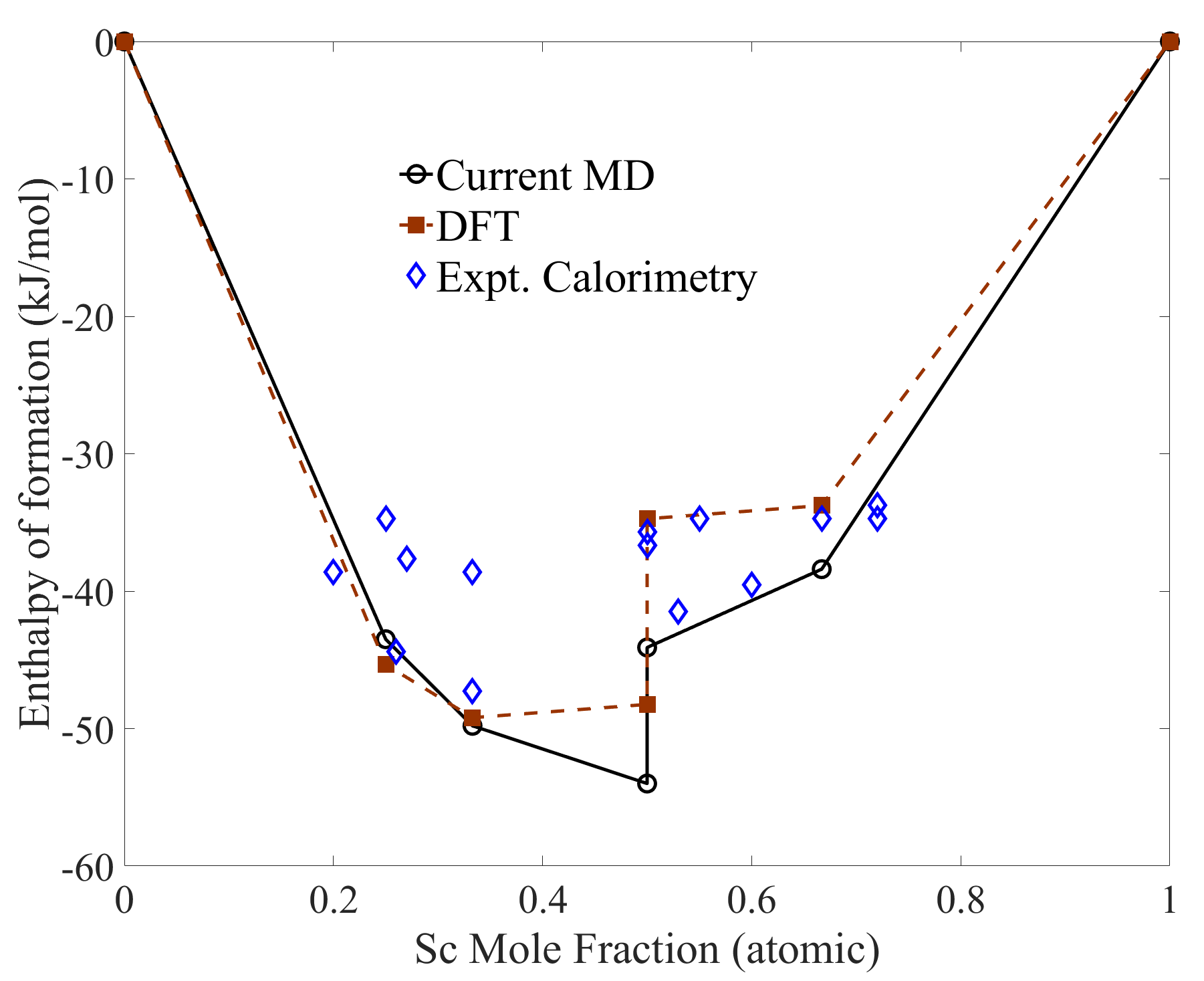}
	\caption{Formation enthalpy ($\Delta H$) of Al--Sc intermetallic compounds as a function of scandium atomic fraction. 
		The black line with open circles represents values computed using the present 2NN--MEAM potential, while the brown dashed line with filled squares corresponds to zero-temperature \textit{ab initio} (VASP) calculations from Asta et~al.~\cite{asta2001structural}. 
		Blue open diamonds denote experimental calorimetry data from Jung \textit{et al.}, Meschel and Kleppa, and Cacciamani \textit{et al.}~\cite{cacciamani1999thermodynamic, jung1991standard, meschel1994standard}. 
		The 2NN--MEAM results closely follow the experimental trend, accurately reproducing the strong stabilization of Al$_3$Sc and Al$_2$Sc.}
	\label{fig:al_sc_formation}
\end{figure}

Figure~\ref{fig:al_sc_formation} compares the formation enthalpies calculated with the present potential against benchmark data, including first-principles Density Functional Theory (DFT) calculations by Asta et~al.~\cite{asta2001structural} and experimental calorimetry measurements. As illustrated in the figure, our 2NN-MEAM results (black line with open circles) are in excellent agreement with both the theoretical (brown dashed line) and experimental (blue diamonds) data. The potential accurately captures the strong negative formation enthalpies, confirming the thermodynamic stability of key intermetallic phases such as the L1$_2$ Al$_3$Sc and C15 Al$_2$Sc compounds, which are crucial for the strengthening mechanisms in these alloys. The close agreement between our calculated formation enthalpies and the established convex hull across the entire scandium concentration range validates the potential's ability to correctly describe the relative stability of the various intermetallic structures.

Beyond the energetic validation, the potential's ability to reproduce structural and mechanical properties was assessed, as summarized in Tables~\ref{tab:al_sc_struct} and \ref{tab:al3sc_props}. Table~\ref{tab:al_sc_struct} presents a broad comparison of the calculated zero-temperature atomic volumes ($V_0$) and bulk moduli ($B_0$) for several Al--Sc compounds against the DFT values from Asta et~al. The agreement is consistently strong across different stoichiometries and crystal structures, from the Al-rich Al$_3$Sc to the Sc-rich AlSc$_2$ phase, confirming the potential's robustness. For pure hcp-Sc, our 2NN-MEAM potential yields an atomic volume of 24.95~\AA$^3$/atom, which is in excellent agreement with the experimental and DFT value of 25.0~\AA$^3$/atom. This confirms the accuracy of the elemental Sc potential, which serves as the foundation for the binary alloy model.

\begin{table}[H]
	\centering
	\caption{Calculated structural and elastic properties for Al--Sc compounds. 
		$V_0$ and $B_0$ denote zero-temperature atomic volumes (\AA$^3$/atom) and bulk moduli, respectively. 
		Current MD results are compared with DFT values of Asta \textit{et al.}$^{\mathrm{c}}$\cite{asta2001structural}. 
		Experimental values at room temperature are shown in parentheses.}
	\label{tab:al_sc_struct}
	
	\renewcommand{\arraystretch}{1.15}
	\begin{tabularx}{\linewidth}{@{}l l 
			>{\centering\arraybackslash}X  
			>{\centering\arraybackslash}X  
			>{\centering\arraybackslash}X  
			>{\centering\arraybackslash}X  
			>{\raggedright\arraybackslash}X@{}} 
		\toprule
		Composition & Structure 
		& $V_0$ (\AA$^3$/atom)$^{\mathrm{c}}$ 
		& $V_0$ (Current MD) 
		& $B_0$ (GPa)$^{\mathrm{c}}$ 
		& $B_0$ (Current MD) 
		& Notes \\
		\midrule
		
		Al & A1 (Cu) 
		& 15.8, 15.7$^{\mathrm{a}}$, 16.0$^{\mathrm{b}}$, 16.6$^{\mathrm{c}}$
		& 16.54\cite{lee2003semiempirical} 
		& 81--83 
		& 79.4 \cite{lee2003semiempirical} 
		& fcc-Al reference \\
		
		Al$_3$Sc & L1$_2$ (Cu$_3$Au) 
		& 16.5, 16.1$^{\mathrm{a}}$, 16.4$^{\mathrm{b}}$, 17.24--17.30$^{\mathrm{c}}$
		& 18.17 
		& 94--96, 99\cite{george1990brittle} 
		& 100.84 
		& Stable intermetallic \\
		
		Al$_2$Sc & C15 (Cu$_2$Mg) 
		& 17.3\,(18.15, 18.28)$^{\mathrm{c}}$ 
		& 17.76 
		& 94 
		& 94.57 
		& Laves phase \\
		
		AlSc & B2 (CsCl) 
		& 17.8--18.3$^{\mathrm{a}}$, 18$^{\mathrm{b}}$, 19.45--20.52$^{\mathrm{c}}$
		& 20.05 
		& 84--87 
		& 99.16
		& Ordered B2 prototype \\
		
		AlSc & B$_f$ (BCr) 
		& 19.2, 18.8$^{\mathrm{a}}$, 19.45$^{\mathrm{c}}$ 
		& 21.24 
		& 76 
		& 74.46 
		& Higher-energy variant \\
		
		AlSc$_2$ & B82 (InNi$_2$) 
		& 20.2--21.3$^{\mathrm{c}}$ 
		& 20.13 
		& 75 
		& 64.89 
		& Sc-rich phase \\
		
		Sc & A3 (Mg) 
		& 22.1--23$^{\mathrm{a}}$, 22.35$^{\mathrm{b}}$, 25.0$^{\mathrm{c}}$ 
		& 24.95 
		& 60--61 
		& 55.29 
		& hcp-Sc reference \\
		\bottomrule
	\end{tabularx}
\end{table}

A more detailed analysis of the primary strengthening precipitate, L1$_2$--Al$_3$Sc, is provided in Table~\ref{tab:al3sc_props}. The calculated lattice parameter of 4.10~\AA\ is in excellent agreement with the experimental value of 4.089~\AA, and the predicted formation enthalpy of $-0.451$~eV/atom matches the experimental value precisely. The calculated elastic properties also show strong agreement with both DFT and experimental data for key metrics. The bulk modulus ($B = 100.8$~GPa) falls well within the experimental range of 92--99~GPa, and the $C_{44}$ shear constant (67.8~GPa) is nearly identical to the measured value of 68~GPa. While the $C_{11}$ and $C_{12}$ constants show some deviation from experimental values, a common trait for empirical potentials, the potential accurately captures the fundamental stiffness and shear resistance of the precipitate. Interestingly, our calculated Poisson’s ratio of 0.33 is considerably higher than the experimental value of 0.18 and the range calculated from Asta et~al.’s DFT data (0.17--0.20). This discrepancy suggests that while the potential correctly models the material's response to pure volume change (bulk modulus) and a specific shear mode ($C_{44}$), the coupling between tensile and compressive strain is overestimated. Nevertheless, the overall robust performance for the most critical thermodynamic and elastic properties confirms that the 2NN-MEAM potential is well-suited for atomistic simulations of solidification and mechanical behavior in the Al--Sc alloy system.

\begin{table}[H]
	\centering
	\captionsetup{justification=raggedright,singlelinecheck=false}
	\caption{Calculated physical and mechanical properties of Al$_3$Sc (L1$_2$). 
		2NN--MEAM results are compared with DFT data from Asta \textit{et al.}$^{\,a}$\cite{asta2001structural} 
		and Chen \textit{et al.}$^{\,b}$\cite{chen2014first}. 
		Experimental values$^{\,c}$ are included for reference. 
		Shear modulus $G$ and Poisson ratio $\nu$ for Asta’s dataset are calculated from $C_{11}$, $C_{12}$, and $C_{44}$ 
		using the Voigt--Reuss--Hill averaging relations for cubic symmetry.}
	\label{tab:al3sc_props}
	
	\renewcommand{\arraystretch}{1.15}
	\begin{tabularx}{\linewidth}{@{}l>{\centering\arraybackslash}X>{\centering\arraybackslash}X>{\centering\arraybackslash}X>{\centering\arraybackslash}X@{}}
		\toprule
		Property & 2NN--MEAM (This work) & Asta \textit{et al.}$^{\,a}$ & Chen \textit{et al.}$^{\,b}$ (GGA/LDA) & Experiment$^{\,c,}$ $^{\,d,}$ $^{\,e}$ \\
		\midrule
		Lattice parameter $a$ (\AA) & 4.10 & 4.05--4.11 & 4.110 / 4.021 & 4.089 \\
		Atomic volume $V_0$ (\AA$^3$/atom) & 18.17 & 16.1--16.7 & 16.5 (GGA), 16.1 (LDA) & 17.24--17.29 \\
		Formation enthalpy $\Delta H$ (eV/atom) & $-$0.451 & $-$0.446 to $-$0.537 & $-$0.497 (GGA), $-$0.523 (LDA) & $-$0.451 \\
		\midrule
		$C_{11}$ (GPa) & 151.9 & 179--191 & 179.5 / 202.3 & 183 \\
		$C_{12}$ (GPa) & 75.3 & 38--46 & 39.2 / 41.9 & 46 \\
		$C_{44}$ (GPa) & 67.8 & 66--82 & 67.6 / 78.2 & 68 \\
		\midrule
		Bulk modulus $B$ (GPa) & 100.8 & 92--96 & 86.6 / 96.3 & 92--99 \\
		Shear modulus $G$ (GPa) & 67.8 & 68--79$^{*}$ & 68.6 / 79.0 & 68 \\
		Young’s modulus $E$ (GPa) & 165 (est.) & 163--185$^{*}$ & 163 / 186 & 166 \\
		Poisson’s ratio $\nu$ & 0.33 & 0.17--0.20$^{*}$ & 0.185 / 0.176 & 0.18 \\
		\bottomrule
	\end{tabularx}
	
	\vspace{0.5em}
	\raggedright
	\footnotesize
	$^{\,a}$~Asta \textit{et al.}, \textit{Phys. Rev. B} \textbf{64}, 094104 (2001)\cite{asta2001structural}.  
	$^{\,b}$~Chen \textit{et al.}, \textit{Comput. Mater. Sci.} \textbf{91}, 165 (2014)\cite{chen2014first}.  
	$^{\,c}$~Experimental data:  
	atomic volume from Villars \& Calvert, \textit{Pearson’s Handbook of Crystallographic Data for Intermetallic Phases} (ASM International, 1991);  
	formation enthalpy from Cacciamani \textit{et al.}, \textit{Intermetallics} \textbf{7} (1999) 101;  
	$^{\,d}$bulk modulus from Hyland \& Stiffler \cite{hyland1991determination}, \textit{Scripta Metall. Mater.} \textbf{25} (1991) 473, and $^{\,e}$George \textit{et al.}, \textit{J. Mater. Res.} \textbf{5} (1990) 1639\cite{george1990brittle}.
	
\end{table}


\subsection{Al-Sc High Temperature Properties}
\begin{figure}[H]
	\centering
	\includegraphics[width=\textwidth]{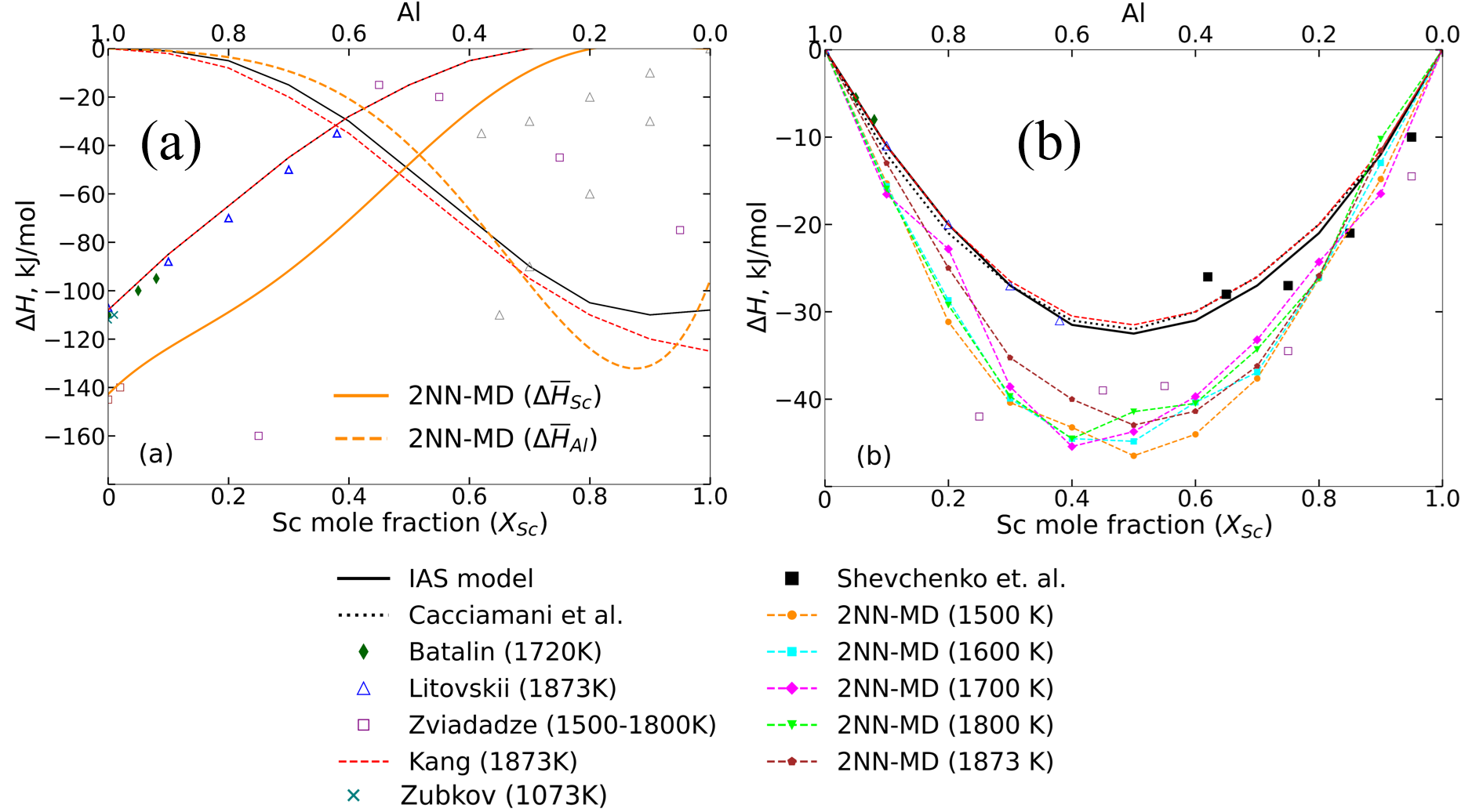}
	\caption{
		Partial (a) and integral (b) mixing enthalpies of liquid Al–Sc alloys compared with available experimental and thermodynamic data. 
		Panel (a) shows the partial molar enthalpies of aluminum ($\Delta \bar{H}_{\mathrm{Al}}$) and scandium ($\Delta \bar{H}_{\mathrm{Sc}}$) obtained from 2NN-MD simulations at 1873~K (orange lines) and compared with the ideal associated solution (IAS) model (black solid and dashed lines). 
		The present results reproduce the strong asymmetry between $\Delta \bar{H}_{\mathrm{Al}}$ and $\Delta \bar{H}_{\mathrm{Sc}}$, with a minimum around $x_{\mathrm{Sc}} \approx 0.5$, consistent with calorimetric measurements and IAS fitting reported by Shevchenko \textit{et~al.}\,\cite{shevchenko2014thermodynamic}. 
		Literature data from Batalin \textit{et~al.}\,(1720~K)\,\cite{Batalin1985}, Litovskii \textit{et~al.}\,(1873~K)\,\cite{litovskii1986enthalpies}, Zviadadze \textit{et~al.}\,(1500–1800~K)\,\cite{zviadadze1982thermodynamics}, and Kang \textit{et~al.}\,(1873~K evaluation)\,\cite{kang2008} are shown for reference. 
		Panel (b) compares the integral enthalpy of mixing ($\Delta H_{\mathrm{mix}}$) predicted by 2NN-MD over 1500–1873~K with the IAS model and prior evaluations by Cacciamani \textit{et~al.}\,\cite{cacciamani1999thermodynamic}, showing excellent agreement near the experimentally observed minimum of $-32.7~\mathrm{kJ/mol}$ at $x_{\mathrm{Sc}} \approx 0.49$. 
		The trends confirm that the 2NN-MD approach accurately reproduces the exothermic nature of mixing in the Al–Sc melt and the strong deviation from ideal-solution behavior observed experimentally. 
	}
	\label{fig:liquid-enthalpy}
\end{figure}

The thermodynamic behavior of liquid Al–Sc alloys was evaluated using the developed 2NN-MEAM potential and benchmarked against calorimetric and model data. 
Figure~\ref{fig:liquid-enthalpy} presents both the partial and integral mixing enthalpies of Al–Sc melts over a wide range of compositions, compared with experimental datasets and the Ideal Associated Solution (IAS) model of Shevchenko~\textit{et~al.}\,\cite{shevchenko2014thermodynamic}. The IAS approach accounts for the formation of distinct associates (\ce{Al3Sc}, \ce{Al2Sc}, \ce{AlSc}, \ce{AlSc2}) in the melt and expresses the partial and integral enthalpies as composition-dependent polynomials:  
\begin{align}
	\Delta \bar{H}_{\mathrm{Sc}} &= (1 - x_{\mathrm{Sc}})^{2}\!\left(-107.7 - 261.7x_{\mathrm{Sc}} + 580.3x_{\mathrm{Sc}}^{2} - 251.9x_{\mathrm{Sc}}^{3}\right), \label{eq:Hsc} \\
	\Delta H_{\mathrm{mix}} &= x_{\mathrm{Sc}}^{2}\!\left(23.1 - 648.5x_{\mathrm{Sc}} + 769.3x_{\mathrm{Sc}}^{2} - 251.9x_{\mathrm{Sc}}^{3}\right), \label{eq:Hmix}
\end{align}
where $x_{\mathrm{Sc}}$ is the scandium mole fraction. These equations quantitatively reproduce the strong exothermic character of mixing, with a minimum $\Delta H_{\mathrm{mix}} = -32.7~\mathrm{kJ/mol}$ at $x_{\mathrm{Sc}} \approx 0.49$.

Panel~(a) of Fig.~\ref{fig:liquid-enthalpy} shows the partial molar enthalpies of Al ($\Delta \bar{H}_{\mathrm{Al}}$) and Sc ($\Delta \bar{H}_{\mathrm{Sc}}$) obtained from the present 2NN-MD simulations at 1873~K (orange lines) in comparison with the IAS model (black solid and dashed lines). 
The results capture the strong asymmetry between $\Delta \bar{H}_{\mathrm{Al}}$ and $\Delta \bar{H}_{\mathrm{Sc}}$, consistent with the calorimetric findings of Shevchenko~\textit{et~al.}\,\cite{shevchenko2014thermodynamic} and earlier studies by Batalin~\textit{et~al.}\,\cite{Batalin1985}, Litovskii~\textit{et~al.}\,\cite{litovskii1986enthalpies}, and Zviadadze~\textit{et~al.}\,\cite{zviadadze1982thermodynamics}. 
\textcolor{red}{A slight discrepancy appears on the Al-rich side for $\Delta \bar{H}_{\mathrm{Sc}}$, where small oscillations arise from the limited number of Sc atoms contributing to the ensemble average in these dilute configurations.} 
\textcolor{red}{These fluctuations are common in MD studies of metallic melts and reflect sampling noise rather than true thermodynamic behavior.} 
\textcolor{red}{Similar behavior was reported in the Mg--Al system analyzed by Kim \textit{et~al.}\,\cite{kim2009atomistic}.} 
\textcolor{red}{These oscillations do not alter the overall thermodynamic trends or the accuracy of the potential.} 
\textcolor{red}{Figure~2(b) compares the temperature-dependent $\Delta H_{\mathrm{mix}}$ predicted by 2NN-MD (1500--1873~K) with the IAS model and prior CALPHAD assessments by Cacciamani \textit{et~al.}\,\cite{cacciamani1999thermodynamic}.} 
\textcolor{red}{The excellent agreement near the experimental minimum confirms that the present interatomic potential reliably captures the exothermic nature of mixing and the strong negative deviation from ideal-solution behavior in high-temperature Al--Sc liquids.}

\subsection{Solidification of Sc}

To provide a theoretical basis for analyzing the solidification behavior, the solid--liquid interfacial free energy ($\sigma_{SL}$) was estimated using Turnbull's empirical rule, which relates the interfacial energy to the enthalpy of fusion and atomic volume at the melting point \cite{turnbull1950}. The relationship is given by:
\begin{equation}
	\sigma_{SL} = C \frac{\Delta h_m}{v_a^{2/3}},
\end{equation}
where $C$ is Turnbull's coefficient, $\Delta h_m$ is the enthalpy of fusion per atom, and $v_a$ is the atomic volume of the solid at the melting temperature. For this calculation, we used the accepted value of $C=0.45$ for close-packed metals. 

The enthalpy of fusion predicted by the 2NN--MEAM potential is $\Delta H_m = 16.1~\text{kJ/mol}$, which corresponds to a per-atom enthalpy of 
\[
\Delta h_m = \frac{16{,}100~\text{J/mol}}{6.022 \times 10^{23}~\text{atoms/mol}} = 2.67 \times 10^{-20}~\text{J/atom}.
\]

The average atomic volume at the melting point ($T_m = 1814~\text{K}$) was obtained directly from our 2NN--MEAM NPT simulations. The final three frames of the $10^6$-step trajectory yielded average cell dimensions of $L_x = 26.48~\text{\AA}$, $L_y = 34.3985~\text{\AA}$, and $L_z = 445.675~\text{\AA}$ for a system of 14{,}400 atoms, giving a total volume of $4.07\times10^5~\text{\AA}^3$ and an average atomic volume of $v_a = 28.25~\text{\AA}^3 = 2.825\times10^{-29}~\text{m}^3$.

Substituting these values into Turnbull's rule yields:
\[
\sigma_{SL} = (0.45) \frac{2.67 \times 10^{-20}~\text{J}}{(2.825 \times 10^{-29}~\text{m}^3)^{2/3}} 
= \frac{1.20 \times 10^{-20}~\text{J}}{9.3 \times 10^{-20}~\text{m}^2}
\approx 0.129~\frac{\text{J}}{\text{m}^2}
= 129~\frac{\text{mJ}}{\text{m}^2}.
\]

This value represents a physically reasonable estimate for scandium. Considering the typical range of Turnbull’s coefficient ($C=0.3$--$0.6$), the corresponding interfacial free energy spans $86$--$173~\text{mJ/m}^2$, providing a useful benchmark for subsequent analysis of nucleation phenomena.

Following the estimation of $\sigma_{SL}$, homogeneous nucleation simulations were conducted on pure liquid Sc using the same 2NN--MEAM potential. We observed spontaneous homogeneous nucleation in the temperature range of 300~K to 850~K, while no nucleation events occurred at or above 900~K. This observation is consistent with the melting point predicted by our simulations ($T_m = 1814~\text{K}$), which closely matches the experimental value, indicating that the free-energy barrier to solidification becomes prohibitively large at higher temperatures. At 900~K, even after 500~ps of simulation time, the system remained completely in the liquid state, highlighting the difficulty of overcoming the nucleation barrier at small undercoolings. In principle, homogeneous nucleation could still occur at such temperatures if the system size were sufficiently increased to allow for larger critical fluctuations or if the simulation were extended to several nanoseconds, providing sufficient time for a rare, large fluctuation to cross the free-energy barrier.

\begin{figure}[H]
	\centering
	\includegraphics[width=\linewidth]{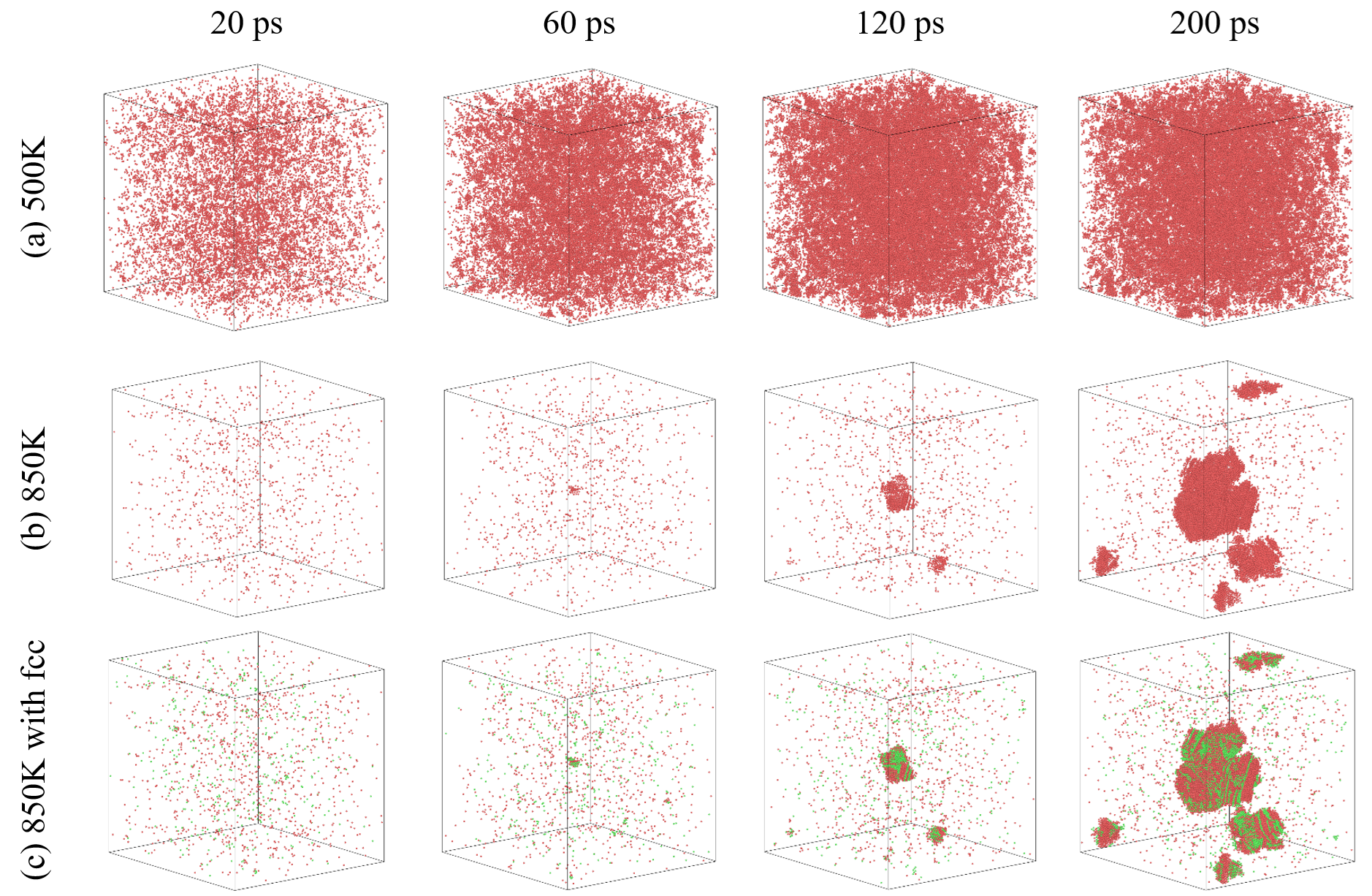}
	\caption{Snapshots of nuclei formation and growth during solidification of Sc for two isothermal processes. Fcc atoms are shown in green, while hcp atoms are shown in red; amorphous atoms are omitted for clarity. (a)~500~K, (b)~850~K, and (c)~850~K with fcc atoms highlighted. For comparison, the nucleation and stacking-fault evolution observed here closely follow similar trends reported for Al by Mahata \textit{et al.}~\cite{mahata2018understanding}.}
	\label{fig:sc_nucleation}
\end{figure}

At undercooled conditions ($T < 850$~K), the nucleation process proceeds through the formation of multiple critical nuclei that subsequently grow and coalesce, similar to the observations of Mahata \textit{et al.} for Al solidification \cite{mahata2018understanding}. As the temperature increases toward the melting point, the time required for the first nucleus to appear increases markedly, reflecting the exponential sensitivity of the nucleation rate to undercooling. At deep undercooling (e.g., 500~K), numerous small nuclei form almost simultaneously throughout the simulation box, leading to rapid crystallization. In contrast, at moderate undercooling (e.g., 850~K), the number of nuclei decreases significantly, and their formation is delayed; however, each nucleus grows substantially larger before impinging on neighboring clusters. This inverse relationship between nucleation density and nucleus size is a hallmark of homogeneous solidification processes, consistent with predictions of classical nucleation theory (CNT). Figure~\ref{fig:sc_nucleation} illustrates the time evolution of solidification in Sc at representative isothermal conditions. At 500~K, spontaneous crystallization is observed within tens of picoseconds, forming a dense population of fine crystalline grains distributed uniformly in the melt. The rapid emergence of multiple nucleation sites leads to simultaneous growth and impingement, producing a polycrystalline network dominated by hexagonal close-packed (hcp) stacking with occasional face-centered cubic (fcc) regions. At 850~K, however, the system exhibits a pronounced induction period: nucleation initiates after nearly 100~ps, followed by slower, anisotropic growth of a few isolated nuclei that evolve into larger, well-faceted crystals over several hundred picoseconds. As shown in Fig.5~(c) of Figure~\ref{fig:sc_nucleation}, the fcc atoms (green) appear primarily along stacking-fault planes or twin boundaries within the predominantly hcp matrix (red), while amorphous liquid atoms are omitted for clarity. This interplay between hcp and fcc coordination during early solidification is characteristic of close-packed metals, where stacking-fault transformations mediate structural rearrangements as the system approaches equilibrium. 

With increasing temperature, the overall nucleation rate decreases, the number of nuclei reduces, and the mean grain size increases, consistent with the temperature dependence of $\Delta G_v$ and $\sigma_{SL}$ in CNT. The trends observed here for Sc mirror those previously reported for Al \cite{mahata2018understanding, mahata2019effects, mahata2019evolution, mahata2019size}, further validating the transferability of the 2NN--MEAM framework for modeling nucleation phenomena in close-packed metallic systems.

\begin{figure}[H]
	\centering
	\includegraphics[width=\linewidth]{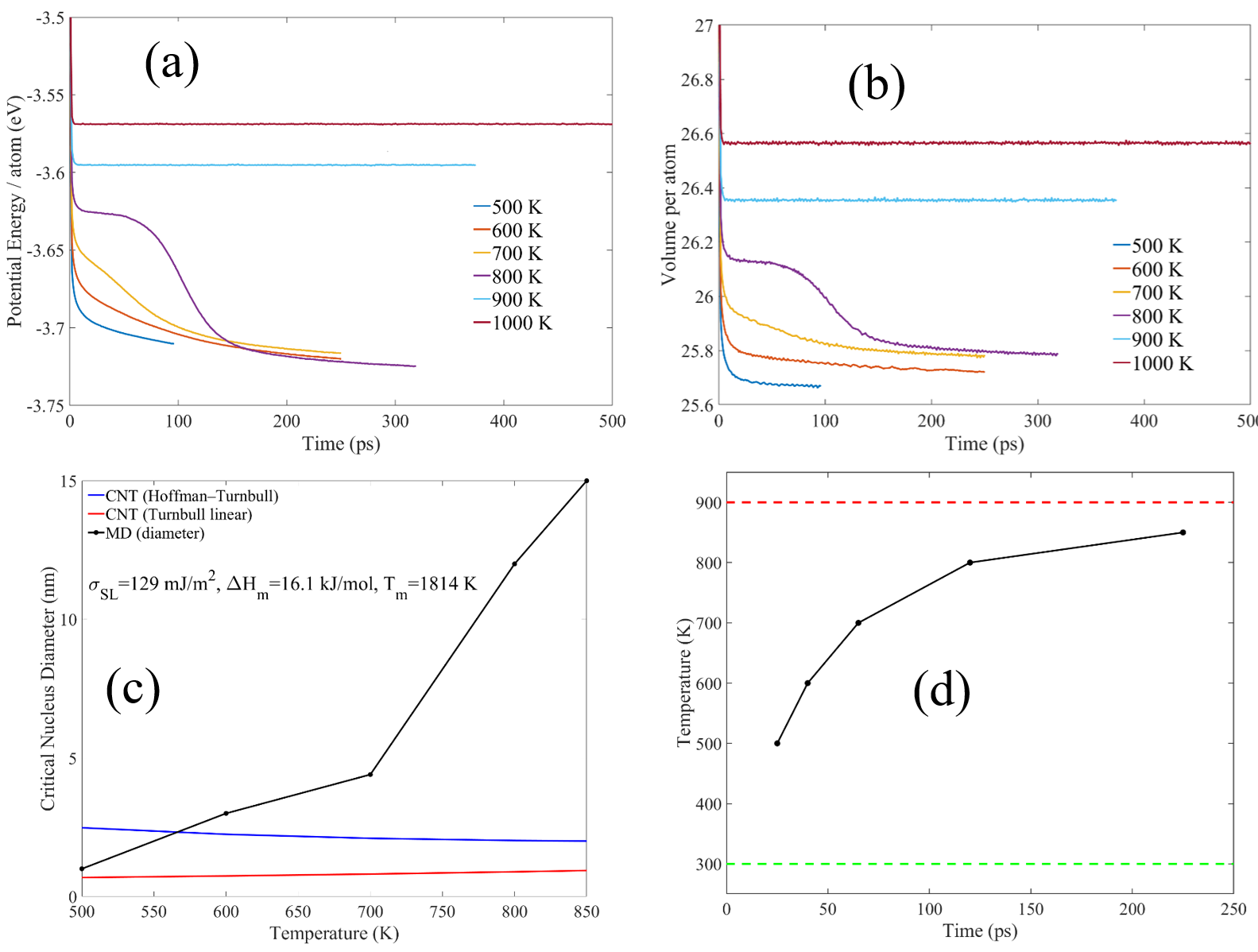}
	\caption{Time evolution for pure Sc at fixed temperature: (a) potential energy per atom (eV/atom) and (b) volume per atom (\AA$^{3}$/atom). Curves at 500--850~K show sharp drops at crystallization, whereas 900--1000~K remain flat, indicating no nucleation within $\sim$0.5~ns. (c)~Critical nucleus \emph{diameter} $d^{\ast}$ from capillarity-based CNT using $V_m(T)$ from panel~(b): Hoffman--Turnbull ($\Delta g_v = \Delta H_m V_m^{-1} T \Delta T / T_m^2$) and Turnbull linear ($\Delta g_v = \Delta H_m V_m^{-1} \Delta T / T_m$), together with MD diameters. (d)~Isothermal TTT summary (time to first nucleus vs.\ temperature); dashed lines mark $T_m / 2$ ($\approx$907~K) and 300~K. Parameters: $\sigma_{SL} = 129$~mJ\,m$^{-2}$, $\Delta H_m = 16.1$~kJ\,mol$^{-1}$, $T_m = 1814$~K.}
	\label{fig:sc_pe_vol}
\end{figure}

The trends in Fig.~\ref{fig:sc_pe_vol} agree with the snapshots in Fig.~\ref{fig:sc_nucleation}. Large undercooling, $\Delta T = T_m - T$, gives a strong bulk driving force $\Delta g_v$ and therefore a small classical barrier, $\Delta G^{\ast} = \tfrac{16\pi}{3}\sigma_{SL}^3 / \Delta g_v^2$. At $500$~K, the liquid crystallizes rapidly and both potential energy and volume per atom drop sharply; at $800$--$850$~K there is a clear induction period followed by a stepwise decrease as a few nuclei form and grow, consistent with the hcp-dominated structures (with fcc along stacking faults) shown in Fig.~\ref{fig:sc_nucleation}. At $900$--$1000$~K, the undercooling is small relative to $T_m \approx 1814$~K, the barrier is large, and no nucleation events are observed within $\sim$0.5~ns, so the traces remain flat. For $T \le 500$~K, the liquid collapses into a metastable solid almost immediately, so transient form--dissolve episodes are not resolved in the thermodynamic signals. These behaviors are consistent with the Turnbull estimate of $\sigma_{SL}$ and our coexistence-based $T_m$ and $\Delta H_m$ (Sec.~3.4).

Following Mahata \textit{et al.}~(2018), we compare the measured critical sizes with two capillarity-based CNT formulations using the \emph{per-temperature} molar volumes $V_m(T)$ taken from the final plateaus in Fig.~\ref{fig:sc_pe_vol}(b). The linear (Turnbull) approximation assumes
\begin{equation}
	\Delta g_v(T) = \frac{\Delta H_m}{V_m(T)} \frac{\Delta T}{T_m},
	\qquad
	r^{\ast}(T) = \frac{2\sigma_{SL}}{\Delta g_v(T)} ,
\end{equation}
while the Hoffman--Turnbull form includes the additional temperature dependence:
\begin{equation}
	\Delta g_v(T) = \frac{\Delta H_m}{V_m(T)} \frac{T\,\Delta T}{T_m^2},
	\qquad
	r^{\ast}(T) = \frac{2\sigma_{SL}}{\Delta g_v(T)} .
\end{equation}
Diameters \( d^{\ast} = 2r^{\ast} \) from both expressions are plotted in Fig.~\ref{fig:sc_pe_vol}(c), together with MD-extracted diameters. The Turnbull curve increases weakly with \( T \) as \( \Delta T \) shrinks, whereas the Hoffman--Turnbull curve decreases slightly for \( T < T_m / 2 \) (here \( T_m / 2 \approx 907~\text{K} \)), as expected from its \( T \Delta T \) dependence.

Fig.5~(d) provides a time--temperature--transformation (TTT) view, showing the characteristic “nose” of the solidification process. The incubation time for the first nucleus increases rapidly with temperature and diverges as $T$ approaches $T_m/2$, consistent with the increasing free-energy barrier and reduced driving force. At deep undercooling ($\leq 600$~K), the short incubation period reflects spontaneous nucleation and fast growth, whereas above $\sim$800~K, the longer delay corresponds to the sluggish formation of a few large clusters that subsequently grow slowly. This kinetic picture complements the thermodynamic trends in Fig.5~(a)--(c).

The absolute CNT sizes lie below the MD diameters at higher temperatures. This deviation is reasonable for several reasons:  
(i)~$\sigma_{SL}$ for Sc is uncertain and anisotropic; Turnbull’s estimate ($\sim$129~mJ\,m$^{-2}$) neglects orientation and curvature (Tolman) corrections;  
(ii)~very small clusters do not obey bulk thermodynamics---$V_m(T)$, $\Delta H_m$, and interfacial structures are cluster-size dependent;  
(iii)~nuclei are mixed hcp/fcc with stacking faults (Fig.~\ref{fig:sc_nucleation}), not the sharp, spherical capillarity objects assumed by CNT;  
(iv)~finite-size stress, residual pressure, and structural-counting definitions in MD bias the apparent $r^{\ast}$ upward; and  
(v)~Sc lacks rigorous experimental $\sigma_{SL}(T)$ and $\Delta\mu(T)$ data, so even though $V_m(T)$ from Fig.~\ref{fig:sc_pe_vol}(b) improves the estimate, the driving force still relies on simplified capillarity approximations.  
These limitations explain the quantitative offset without affecting the qualitative picture: as undercooling decreases, $\Delta g_v$ falls, barriers rise, incubation times grow, and nucleation becomes increasingly rare.

\subsection{Homogeneous Nucleation during Solidification of Al--Sc}

\begin{figure}[H]
	\centering
	\includegraphics[width=0.98\linewidth]{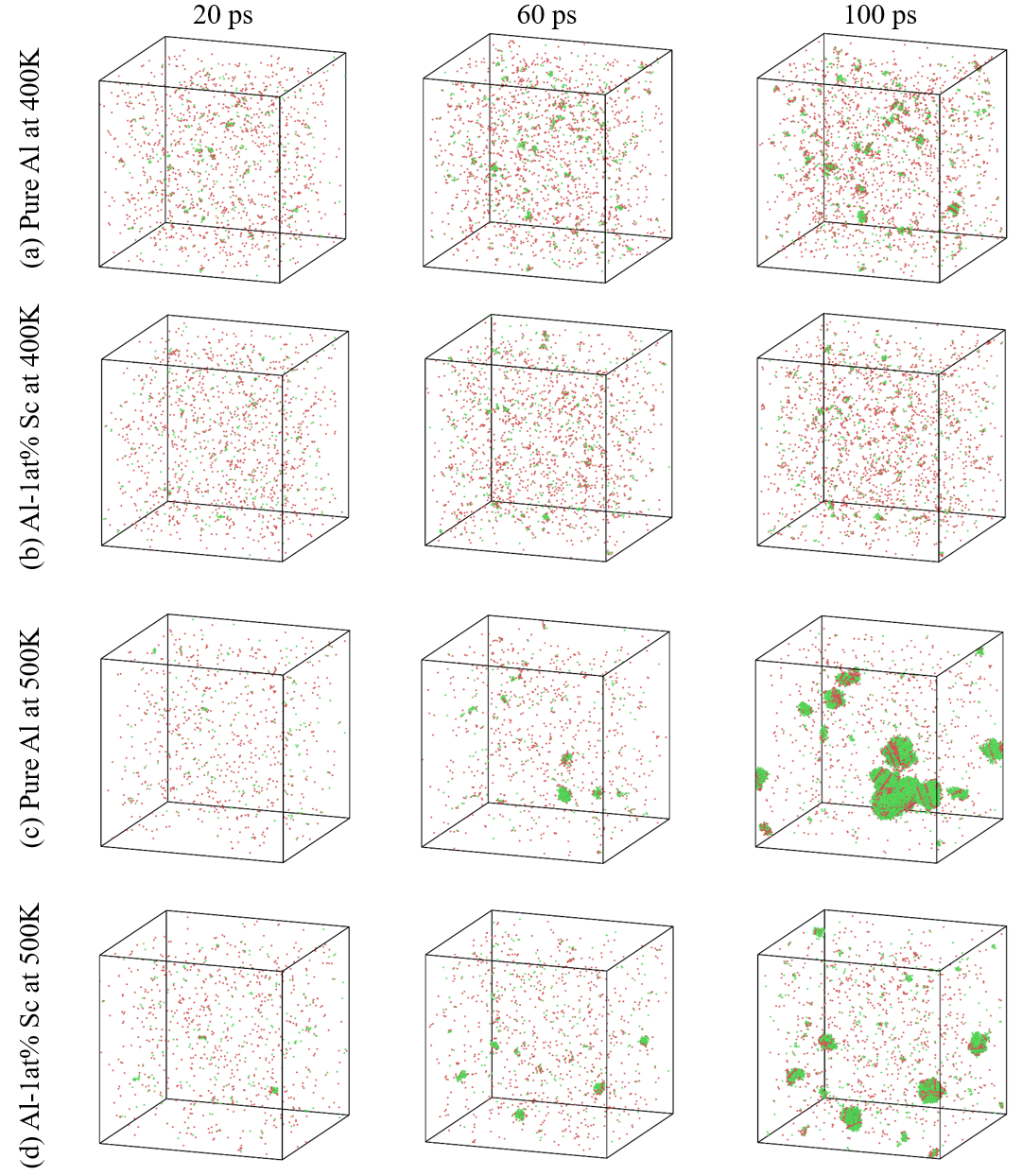}
	\caption{Time-resolved snapshots of homogeneous nucleation in pure Al and Al--1~at.\% Sc at two temperatures. Columns correspond to 20, 60, and 100~ps. Rows (a)--(d) label composition and temperature. fcc atoms are highlighted; amorphous atoms are omitted for clarity. At equal absolute $T$, pure Al forms larger and more numerous clusters than Al--1~at.\% Sc by 100~ps, consistent with a larger driving force and the absence of solute drag.}
	\label{fig:Al-Sc-Nucleation}
\end{figure}

Figure~\ref{fig:Al-Sc-Nucleation} compares homogeneous nucleation in pure Al and in Al--1~at.\% Sc at two isothermal conditions. Each row shows the same system viewed at 20, 60, and 100~ps, while columns (a)--(d) correspond to the composition–temperature pairs annotated at left. In the Al boxes, compact fcc clusters appear by 60~ps and coarsen rapidly by 100~ps, consistent with a strong bulk driving force and fast interface mobility at these undercoolings. In the Al--1~at.\% Sc boxes, only small clusters are visible at 60~ps and the total solid fraction remains lower at 100~ps. The snapshots therefore indicate a longer incubation time and a smaller growth rate for the Sc-bearing melt at the same absolute temperature. This trend agrees with the thermodynamic picture established earlier for close-packed metals: the addition of Sc stabilizes the liquid through favorable Al–Sc association, which reduces the magnitude of the volumetric driving force $\Delta g_v$ at fixed $T$, and the interface feels solute drag as Sc redistributes ahead of the advancing front. In the capillarity framework used in Sec.~3.4, the nucleation barrier $\Delta G^{\ast}=\tfrac{16\pi}{3}\sigma_{SL}^3/\Delta g_v^2$ and the critical radius $r^{\ast}=2\sigma_{SL}/\Delta g_v$ both increase when $|\Delta g_v|$ decreases, so nuclei appear later and grow more slowly in the alloy than in pure Al at the same temperature. Moreover, since Sc raises the alloy liquidus relative to pure Al, equal absolute temperatures correspond to smaller undercoolings $\Delta T=T_{\mathrm{liq}}(x_{\mathrm{Sc}})-T$ in the alloy, which further lowers $|\Delta g_v|$ and depresses both the nucleation rate and the early growth velocity. These qualitative features mirror the Sc-only solidification trends established in Sec.~3.4 and validate that the present potential yields the expected composition dependence of nucleation kinetics. :contentReference[oaicite:0]{index=0}

The structural signatures in Fig.~\ref{fig:Al-Sc-Nucleation} also reflect the chemistry of ordering. In the alloy, early supercritical clusters exhibit Cu$_3$Au-type L1$_2$ environments, where Sc atoms occupy B-sites with twelve Al nearest neighbors and Al occupies A-sites with mixed Al/Sc coordination. Because the L1$_2$ motif is geometrically fcc, polyhedral template matching classifies these regions as fcc while a chemical-order analysis resolves the sublattices. The gradual enrichment of L1$_2$-like neighborhoods within the growing clusters, coupled with the slower coarsening at a given $T$, is consistent with solute partitioning and the additional diffusion needed to achieve chemical order during solidification of Al--Sc. Together, the time-ordered panels (a)--(d) show that the alloy forms fewer and smaller clusters than pure Al by 100~ps at both temperatures, and that chemical ordering emerges as part of the same crystallization process.

\begin{figure}[H]
	\centering
	\includegraphics[width=0.78\linewidth]{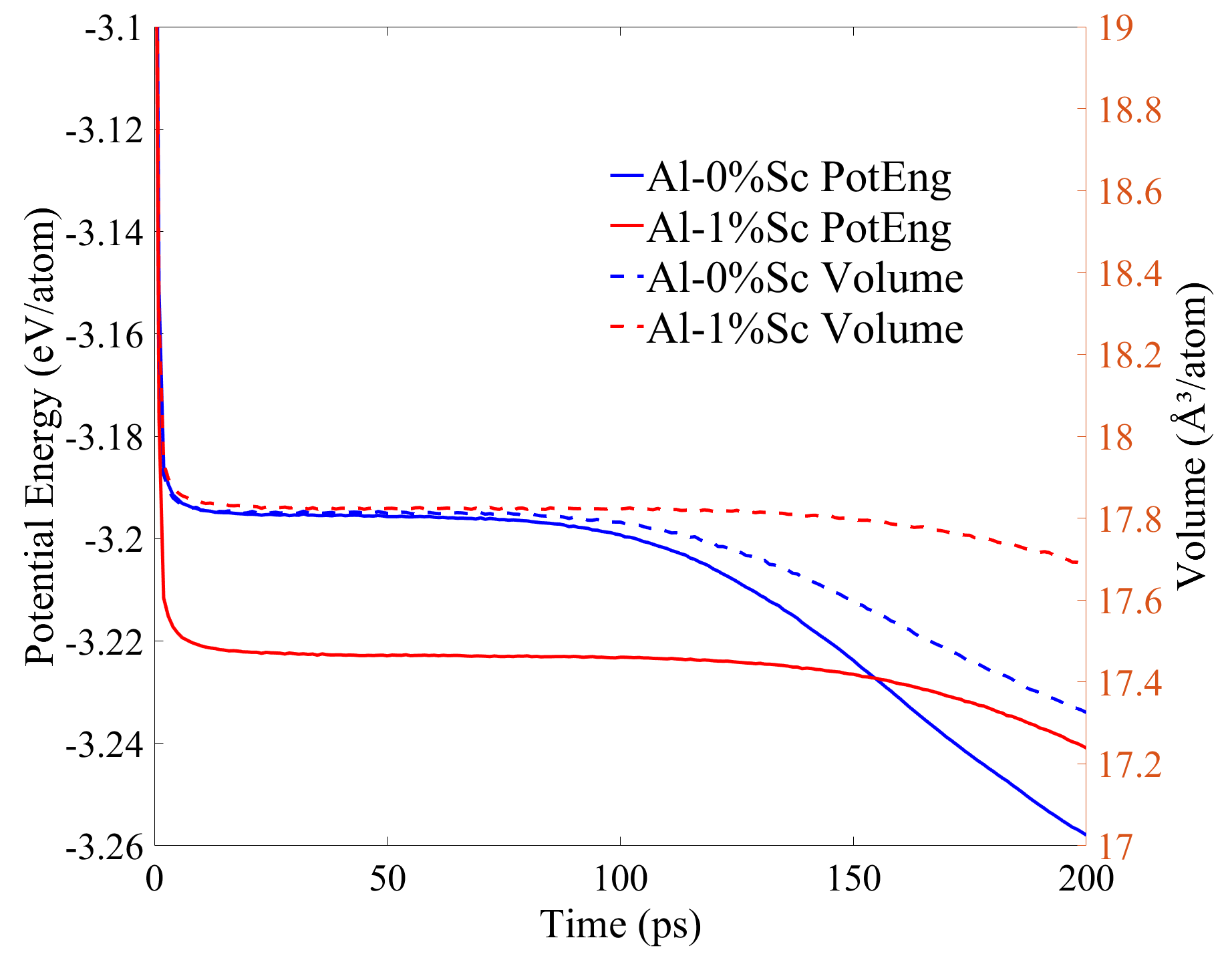}
	\caption{Potential energy per atom (left axis, solid lines) and volume per atom (right axis, dashed lines) during isothermal holds for pure Al (blue) and Al--1~at.\% Sc (red). Pure Al shows a distinct late-time energy drop and volumetric contraction associated with crystallization, whereas the alloy exhibits a smaller secondary energy change and modest volume decrease over the same window, indicating a lower solid fraction and slower growth at equal absolute temperature.}
	\label{fig:Al-Sc-pe-volume}
\end{figure}

The thermodynamic signatures of these events are summarized in Fig.~\ref{fig:Al-Sc-pe-volume}. The potential energy per atom in pure Al (blue solid line) decays modestly during early liquid relaxation, then exhibits a pronounced additional drop after $\sim$80--100~ps as multiple fcc clusters become supercritical and grow. The volume per atom (blue dashed line, right axis) decreases in concert, indicating densification associated with crystallization. In Al--1~at.\% Sc, the potential energy (red solid line) relaxes rapidly at early times to a lower baseline, reflecting strong Al–Sc association in the liquid and the formation of small L1$_2$-like embryos, and then changes only weakly thereafter over the same time window. The corresponding volume trace (red dashed line) remains higher than the pure-Al curve over the entire interval and shows only a slight late-time decrease. The combination of a smaller secondary energy drop and a much smaller volumetric contraction demonstrates that, at equal absolute temperature, the alloy proceeds more slowly toward a percolated solid fraction. This contrast is exactly what is expected from classical nucleation theory when the alloy experiences a smaller undercooling and additional kinetic resistance at the interface. In the notation of Sec.~3.4, the weaker pure-Al-like energy–volume “kink” in the alloy reflects a smaller $|\Delta g_v|$ and a larger $r^{\ast}$, while the separation between the dashed curves quantifies the smaller crystallized fraction over the first 200~ps.

Finally, it is important to interpret the composition–temperature comparison in terms of undercooling. Because $T_{\mathrm{liq}}$ increases with $x_{\mathrm{Sc}}$, equal absolute $T$ does not imply equal $\Delta T$. When the data are analyzed at fixed $\Delta T=T_{\mathrm{liq}}(x_{\mathrm{Sc}})-T$ rather than fixed $T$, the relative lag in the alloy diminishes, yet a residual slowdown remains due to solute drag and the time required to establish L1$_2$ order. Thus the figures together provide a coherent picture: the potential captures chemically ordered Al$_3$Sc formation during solidification, and it reproduces the expected reduction in nucleation rate and growth kinetics for dilute Al--Sc relative to pure Al at the same temperature, with energy and volume traces that are consistent with the capillarity relations and the driving-force estimates detailed earlier.

\subsection{Intermetallic Formation during Al--Sc Solidification}

\begin{figure}[t]
	\centering
	\includegraphics[width=0.95\linewidth]{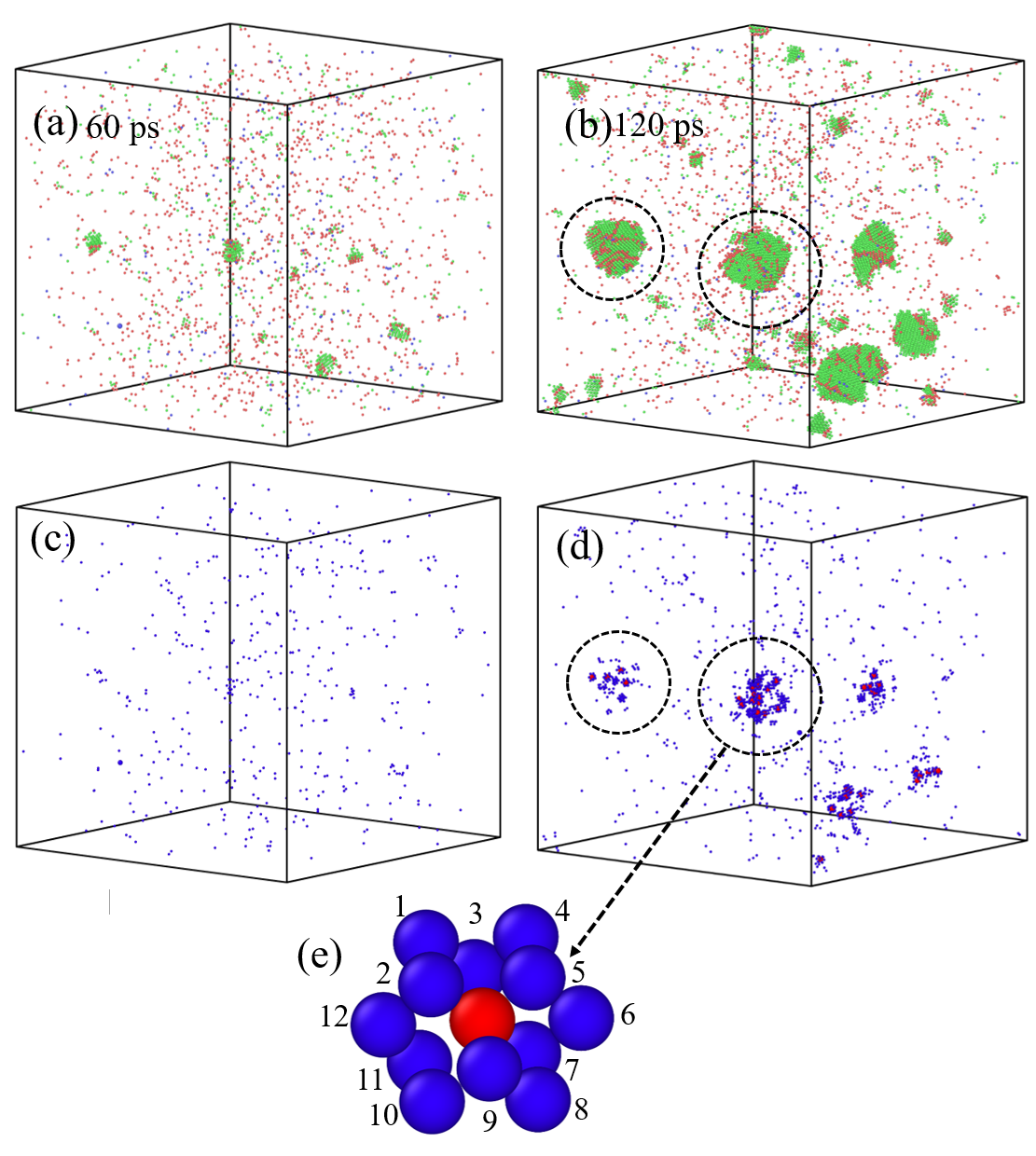}
	\caption{Atomic configurations illustrating intermetallic ordering during solidification of Al--1~at.\% Sc at 500~K. 
		(a) 60~ps: scattered fcc embryos in the undercooled melt. 
		(b) 120~ps: larger Sc-centered clusters showing early ordering. 
		(c)--(d) corresponding L1$_2$-ordered atoms identified by the chemical-order filter, highlighting the formation and growth of Al$_3$Sc domains. 
		(e) The local coordination environment of an L1$_2$ unit, showing a Sc atom (red) surrounded by twelve Al atoms (blue) in the characteristic cuboctahedral geometry.}
	\label{fig:Fcc-L12png}
\end{figure}

Figure~\ref{fig:Fcc-L12png} shows the atomic-scale evolution of ordered Al$_3$Sc during solidification of an Al--1~at.\% Sc alloy. The sequence in Figs.~\ref{fig:Fcc-L12png}(a)--(b) corresponds to structural snapshots extracted at 60 and 120~ps, respectively, from the isothermal simulation at 500~K. The fcc atoms are shown in green and the liquid atoms in red, while dashed circles indicate regions where Sc-centered clusters are forming within the Al matrix. At 60~ps (Fig.~\ref{fig:Fcc-L12png}a), the simulation reveals several small, transiently ordered clusters nucleating from the undercooled melt. By 120~ps (Fig.~\ref{fig:Fcc-L12png}b), these clusters have grown and merged into larger ordered domains, exhibiting the early stages of Al$_3$Sc-type chemical ordering. The densification and ordering are localized around Sc-rich regions, consistent with the strong affinity between Al and Sc established in the mixing enthalpy results presented earlier in Sec.~3.3.

The atomic configurations in Figs.~\ref{fig:Fcc-L12png}(c)--(d) isolate only the atoms identified as chemically ordered L1$_2$ by the polyhedral template matching and ordering-type filters. At 60~ps (Fig.~\ref{fig:Fcc-L12png}c), a small number of L1$_2$ fragments are already present, and by 120~ps (Fig.~\ref{fig:Fcc-L12png}d) these fragments evolve into distinct, Sc-centered ordered clusters. The dashed circles highlight regions where the L1$_2$ order parameter rises sharply, marking the nucleation of Al$_3$Sc embryos within the evolving fcc matrix. The inset (Fig.~\ref{fig:Fcc-L12png}e) magnifies the fundamental coordination geometry, showing one Sc atom (red) surrounded by twelve Al atoms (blue) in a cuboctahedral configuration. This 1:12 coordination represents the characteristic B-site environment of the L1$_2$ (Cu$_3$Au-type) structure, confirming that the simulation captures the correct atomic topology of Al$_3$Sc. 

The emergence of these ordered clusters demonstrates that the potential reproduces the thermodynamic tendency for Sc to occupy the cube-corner sites and Al to occupy the face-center sites in the L1$_2$ lattice. The ordering is driven by the large negative formation enthalpy of Al$_3$Sc ($\Delta H_f = -0.45$~eV/atom; see Table~5), which provides a strong energetic preference for Al--Sc chemical ordering even at very low Sc concentrations. The structural evolution observed here therefore represents the atomistic onset of intermetallic phase formation directly from the undercooled liquid, preceding any long-range precipitation or coarsening.

The observed delay in ordering relative to fcc nucleation reflects the additional kinetic requirements of solute diffusion and site-specific occupancy necessary to achieve full L1$_2$ symmetry. At a given temperature, the number density and growth rate of ordered clusters are lower than those of pure Al nuclei, consistent with a smaller effective undercooling $\Delta T = T_\mathrm{liq}(x_\mathrm{Sc}) - T$ and with the solute-drag effect discussed in Sec.~3.5. These results confirm that during solidification, scandium atoms promote the stabilization of coherent Al$_3$Sc nuclei that subsequently serve as heterogeneous sites for fcc-Al growth, thus linking atomic ordering directly to solidification kinetics. The nucleation of ordered domains therefore constitutes the earliest stage of the microstructural evolution responsible for the exceptional strengthening and thermal stability of Al--Sc alloys. 

In summary, Figs.~\ref{fig:Fcc-L12png}(a)--(d) illustrate the time-resolved formation and growth of Sc-centered ordered clusters, while Fig.~\ref{fig:Fcc-L12png}(e) provides direct evidence of the characteristic 12-fold Al coordination defining the L1$_2$ structure. The simulation thus captures the full pathway from an initially homogeneous liquid to an ordered Al$_3$Sc configuration at the atomic scale. More quantitative studies of nucleation kinetics, critical nucleus size, and the influence of scandium concentration on solidification behavior will be pursued in future publications.

\section{Conclusions}

This work establishes a physically consistent 2NN--MEAM potential for scandium and Al--Sc alloys that accurately captures cohesive, thermodynamic, and solidification behavior from first principles to microstructure. The potential combines the reliability of the elemental 2NN--MEAM formalism with a new Al--Sc parameterization referenced to the L1$_2$--Al$_3$Sc structure. For elemental Sc, the calculated cohesive energy ($E_\mathrm{c}=-3.89$~eV/atom), lattice parameters ($a=3.30$~\AA, $c/a=1.61$), and bulk modulus ($B=55$~GPa) agree with experimental and DFT benchmarks within 1--5\%. Two-phase coexistence simulations yield a melting point of $T_\mathrm{m}=1814$~K and an enthalpy of fusion of $\Delta H_\mathrm{m}=16.1$~kJ~mol$^{-1}$, while the estimated solid--liquid interfacial free energy ($\sigma_{SL}=129$~mJ~m$^{-2}$) provides a quantitative foundation for interpreting nucleation kinetics. The linear thermal expansion coefficient ($10.2\times10^{-6}$~K$^{-1}$) and specific heat ($C_p=25.1$~J~mol$^{-1}$K$^{-1}$) confirm proper anharmonic response at finite temperatures.  

For the Al--Sc binary system, the potential reproduces the strongly negative formation enthalpy of L1$_2$--Al$_3$Sc ($\Delta H_f=-0.451$~eV/atom) and the correct stability hierarchy among Al$_3$Sc, Al$_2$Sc, and AlSc phases. The computed lattice parameter ($a=4.10$~\AA) and elastic constants ($C_{11}=151.9$~GPa, $C_{12}=75.3$~GPa, $C_{44}=67.8$~GPa) closely match experimental values, yielding a bulk modulus of 100.8~GPa, within the experimental range of 92--99~GPa. High-temperature liquid simulations reproduce the measured exothermic mixing enthalpy ($\Delta H_\mathrm{mix}\approx-32$~kJ~mol$^{-1}$ near $x_\mathrm{Sc}=0.5$) and the strong deviation from ideal-solution behavior predicted by the Ideal Associated Solution and CALPHAD models. These thermodynamic results confirm that the potential accurately captures the Al--Sc chemical affinity responsible for ordered compound formation and grain refinement.

Large-scale molecular dynamics simulations of solidification demonstrate that the potential also reproduces the expected kinetic and thermodynamic scaling of homogeneous nucleation. For pure Al, fcc nuclei form rapidly with clear incubation and growth regimes, whereas in Al--1~at.\%~Sc, the nucleation rate is lower and the onset delayed at the same absolute temperature. This behavior arises from a smaller effective undercooling $\Delta T=T_\mathrm{liq}(x_\mathrm{Sc})-T$ and the solute-drag penalty associated with Sc partitioning at the solid--liquid interface. The simulations further resolve Sc-centered L1$_2$--Al$_3$Sc embryos within fcc clusters, where Sc atoms occupy cube-corner (B) sites surrounded by twelve Al neighbors. The evolution of potential energy and atomic volume traces exhibits synchronized signatures: pure Al shows a sharp energy drop and volume contraction at crystallization, while the Al--Sc alloy displays a smaller, slower transition, consistent with reduced driving force $|\Delta g_v|$ and larger critical radius $r^* = 2\sigma_{SL}/\Delta g_v$ derived from classical nucleation theory.

Together, these results demonstrate that the developed potential unifies cohesive energetics, liquid thermodynamics, and solidification pathways in a single transferable framework. It provides a quantitative tool to investigate microalloying effects, interface kinetics, and ordered precipitate formation in Al--Sc and related systems. The present simulations were limited to $\sim$1~million atoms and sub-nanosecond timescales; extending to multi-million-atom domains and microsecond trajectories will enable direct determination of nucleation rates, interfacial anisotropy, and time--temperature--transformation behavior. These larger-scale studies, to be pursued in future work, will bridge atomistic thermodynamics to experimentally observable grain structures and strengthen the predictive capability of MEAM-based alloy modeling for advanced manufacturing applications.

\section*{Acknowledgments}
This work was supported by the National Science Foundation through ACCESS supercomputing allocations MAT250103 and MAT240094. Additional computational resources were provided by Argonne National Laboratory under the Director’s Discretionary allocation for the project \textit{AIAlloyLW}. In addition to this, computational resources for this work were provided by a National Science Foundation MRI Award granted to Wilkes University (Award 1920129)

\section*{Author contribution}
Avik Mahata: Conceptualization, Methodology, Software, Formal analysis, Writing---Original draft preparation.

\section*{Data availability}
All data supporting the findings of this study, including the interatomic potential files and LAMMPS input scripts for Sc and Al–Sc systems, are openly available at:
\url{https://github.com/mahata-lab/Al-Sc-Interatomic-Potential}

\section*{Supplementary materials}
The MEAM potential and library files are included as supplementary materials and are also openly available in the repository linked in the \textit{Data availability} section.

\bibliography{references}           

\end{document}